\documentclass[a4paper,onecolumn,10.7pt,accepted=2026-06-05]{quantumarticle}
\pdfoutput=1

\usepackage{amsmath}
\usepackage{amssymb}
\usepackage{bbm}
\usepackage{amsthm, thm-restate}

\usepackage{graphicx}

\usepackage{xcolor, soul, cancel}
\usepackage{ulem}
\usepackage{wrapfig}
\input{insbox}

\usepackage[linktocpage]{hyperref}
\usepackage[capitalise]{cleveref}
\usepackage{cite}

\crefname{figure}{Figure}{figures}

\usepackage{algpseudocode}
\usepackage{enumerate}

\usepackage[]{mdframed}

\newcounter{lemmaN}

\newcounter{lemmaA}

\newcounter{thoughtexperiment}
\newcounter{restriction}
\renewcommand{\therestriction}{\Alph{restriction}}

\newcounter{colN}

\newtheorem{definition}{Definition}
\newtheorem{theorem}{Theorem}

\newtheorem{claim}{Claim}

\makeatletter
\newcommand*{\balancecolsandclearpage}{
  \close@column@grid
  \clearpage
  \twocolumngrid
}
\makeatother

\makeatletter
\def\tocdepth@fullmunge{
\let\l@section@saved\l@section
\let\l@section\@gobble@tw@
\let\l@subsection@saved\l@subsection
\let\l@subsection\@gobble@tw@
}
\def\tocdepth@fullrestore{
\let\l@section\l@section@saved
\let\l@subsection\l@subsection@saved
}
\newcommand{\hidetoc}[0]{\addtocontents{toc}{\string\tocdepth@fullmunge}}
\newcommand{\restoretoc}[0]{\addtocontents{toc}{\string\tocdepth@fullrestore}}
\makeatother

\newcommand{\IQOQI}{Institute for Quantum Optics and Quantum Information, Austrian Academy of Sciences, Boltzmanngasse 3, A-1090 Vienna, Austria}
\newcommand{\Peri}{Perimeter Institute for Theoretical Physics, 31 Caroline Street North, Waterloo, ON N2L 2Y5, Canada}
\newcommand{\VCQ}{Vienna Center for Quantum Science and Technology (VCQ), Faculty of Physics, University of Vienna, Boltzmanngasse 5, A-1090 
Vienna, Austria}

\makeatletter
\renewcommand*\l@subsection{\@dottedtocline{1}{1.5em}{2em}}
\makeatother

\begin{document}

\title{On the significance of Wigner's Friend in contexts beyond quantum foundations}

\author{Caroline L.\ Jones}
\email{CarolineLouise.Jones@oeaw.ac.at}
\affiliation{\IQOQI{}}
\affiliation{\VCQ{}}
\author{Markus P.\ M\"uller}
\email{Markus.Mueller@oeaw.ac.at}
\affiliation{\IQOQI{}}
\affiliation{\VCQ{}}
\affiliation{\Peri{}}

\date{19 June 2026}

\begin{abstract}
There has been a surge of recent interest in the Wigner's Friend paradox, sparking several novel thought experiments and no-go theorems. The main narrative has been that Wigner's Friend highlights a counterintuitive feature that is unique to quantum theory, and which is closely related to the quantum measurement problem. Here, we challenge this view. We argue that the gist of the Wigner's Friend paradox can be reproduced without assuming quantum physics, and that it underlies a much broader class of enigmas in the foundations of physics and philosophy. To show this, we first consider several recently proposed Extended Wigner's Friend scenarios, and demonstrate that some of their implications for the absoluteness of observations can be reproduced by classical thought experiments that involve the duplication of agents. Importantly, some of these classical scenarios are technologically much easier to implement than their quantum counterparts. Then, we argue that the essential structural ingredient of all these scenarios is a feature that we call ``Restriction A'': that a physical theory cannot give us a joint probabilistic description of the observations of all agents. Finally, we argue that this difficulty is at the core of other puzzles in the foundations of physics and philosophy, and demonstrate this explicitly for cosmology's Boltzmann brain problem. Our analysis suggests that Wigner's Friend should be studied in a larger context, addressing a frontier of human knowledge beyond quantum foundations: to obtain reliable predictions for experiments in which these predictions can be privately but not intersubjectively verified.
\end{abstract}

\maketitle

\tableofcontents

\section{Introduction}

In 1961, Eugene Wigner introduced his now famous thought experiment~\cite{Wigner}, illustrating an important subtlety of what is usually called the quantum \textit{measurement problem}~\cite{Maudlin,BruknerMeasurement}. A friend (F) observes the outcome of a measurement on a quantum system S, either seeing a flash or not. How should a superobserver (say, Wigner) describe the situation? On the one hand, both S and F are quantum systems, and then it follows from linearity that the system and the Friend should be described by the entangled state FS. On the other hand, regarding the Friend, ``the question whether he did or did not see the flash was already decided in his mind, before [Wigner] asked him''~\cite{Wigner}. Thus, it seems as if the correct description of the quantum state after F's measurement would be an updated quantum state that contains only a single term and not a superposition of alternatives. Indeed, as proposed by Deutsch~\cite{Deutsch} (see also~\cite{BruknerMeasurement}), the Friend might even communicate with Wigner and send him a message such as \textit{``I have seen a definite outcome''}. Providing the Friend does not communicate \textit{which} outcome they have seen, this does not alter any subsequent statistical predictions. Therefore this information does not change Wigner's state assignment, yet at the same time challenges his description of an indefinite Friend.

Recently, there has been a resurgence of interest in the Wigner's Friend (WF) thought experiment and its potential implications, via so-called Extended Wigner's Friend (EWF) scenarios. In 2016, Frauchiger and Renner~\cite{Frauchiger} introduced a version involving four agents, showing that it is in general inconsistent for such agents to reason indirectly by pooling each others' predictions, even if those predictions involve only statements of probability zero or one. Brukner~\cite{Brukner} analysed a further multi-agent version of Wigner's Friend by combining Wigner's setup with a Bell scenario, describing his result as a ``no-go theorem for observer-independent facts''. Building on this work, Bong et al.~\cite{Bong} derive a similar conclusion based solely on \textit{actually observed events}, thus demarcating the captured phenomenon from that of Kochen-Specker contextuality~\cite{Kochen}. Many further recent publications have considered aspects of the Wigner's Friend scenario and potential resolutions of its apparently paradoxical predictions and interpretations~\cite{Sudbery,HealeyWF,Baumann3,Baumann1,DeBrota,Relano,Guerin,Baumann2,Leegwater,Haddara,Xu,Walleghem2024}.

It is typically understood that Wigner's Friend is an enigma specific to {quantum theory}: after all, the thought experiments mentioned above all rely on characteristic quantum phenomena, such as superposition or entanglement, or the violation of Bell-type inequalities. However, in this paper we ask whether the consequences for the involved agents and their observations could also be achieved in different settings, \textit{without} quantum theory. In particular, a substantial part of the community take EWF arguments to indicate that quantum theory is incompatible with certain notions of ``absoluteness'' of observations, and of how the observations of different agents relate to one another. We ask though whether this is a uniquely quantum feature, or whether it could be true of other physical theories -- even of classical probability theory. To do so, we will study several thought experiments that are not intrinsically quantum, but that posit similar microscopic interventions over agents: the classical duplication of agents (in several versions), and cosmology's Boltzmann brain problem. We show that important structural features of EWF experiments are reproduced by these scenarios.

More concretely, we argue that there is a common structural core to all these thought experiments and others: a feature that we call ``Restriction \ref{r:A}''. In a nutshell, Restriction \ref{r:A} says that our physical theories cannot always give us a probabilistic description of the observations of all agents. Conditional on assuming the validity of the other two metaphysical statements, the results by Bong et al.\ imply a violation of Absoluteness of Observed Events, which we argue is a (particularly dramatic) instance of Restriction \ref{r:A}. However, we argue that there are also important examples of Restriction \ref{r:A} beyond quantum physics, and, in particular, in classical scenarios.

Before summarising our conclusions, let us clarify the motivation for this work further with an example. Consider the following analogy between our notion of Restriction \ref{r:A} and the concept of \textit{correlation}. In many popular-scientific accounts of quantum theory, the notion of entanglement is explained in an overly simplified and hence incorrect way, similarly as follows: \textit{Given the two electrons in a singlet state, finding that one electron has spin-up allows us to infer} instantaneously \textit{that the other electron has spin-down --- a very puzzling and counterintuitive feature of ``spooky action at a distance''. Here is how it follows from the mathematics of the state vector:...} However, the feature that is described here is not entanglement, but correlation. One can certainly use quantum theory, and its prediction of entanglement, to  \textit{derive} that the phenomenon of correlation appears in physics, but by no means is entanglement \textit{necessary} to obtain physical situations where correlation applies. In particular, the phenomenon just described can be obtained with a pair of shoes that are randomly packaged into two boxes and sent to two agents. Correlation is a phenomenon of probability theory, and therefore of immense importance in classical physics and everyday life. What is specifically quantum is not the phenomenon of correlation, but the specific \textit{form} of correlations that quantum theory admits (namely, ones that violate Bell-type inequalities). Similarly, we argue here that the difficulty of describing all agents with a single joint probability distribution is not specific to quantum theory, even though some of its implementation details are (such as the violation of Local Friendliness inequalities). That is, while quantum EWF scenarios demonstrate (under further assumptions) instances of Restriction \ref{r:A}, we show that this feature appears in classical physics too.

Our paper is divided into two main sections:
\begin{itemize}
    \item In Section \ref{SecWFQC}, we look closely at the metaphysical assumptions of EWF scenarios. We argue that some of the assumptions for EWF scenarios, regarding the \textit{absoluteness} of observations (Subsection \ref{classical_experiments}) or the \textit{consistency} of agents' beliefs (Subsection \ref{reasoning}), also fail to hold in classical thought experiments. In particular, we present two thought experiments, which involve the classical duplication of agents. We argue that, due to the absence of the traditional notion of personal identity, the assumptions are untenable even without any quantum ingredients.
    \item In Section \ref{SecContext}, we formulate a new, structural property of physical theories, which we call Restriction \ref{r:A}. This is a mathematical statement about the impossibility of defining a probability distribution for the observations of agents in certain situations. We argue that Restriction \ref{r:A} is the structural core of EWF scenarios (notably including Local Friendliness and the Frauchiger-Renner thought experiment), as well as of other, classical puzzles in physics and philosophy (such as the Boltzmann brain problem).
    \item From this, we conclude (see Section~\ref{SecConclusions}) that Wigner's Friend has significance far beyond quantum physics. EWF scenarios demonstrate in a particularly dramatic way that our current physical theories cannot always tell agents what they should expect to observe (Restriction~\ref{r:A} for $N=1$ observer). Moreover, even when they do, those predictions cannot be obtained as marginals of a joint prediction for the observations of \textit{all} agents (Restriction~\ref{r:A} for $N\geq 2$ agents). We conclude that Wigner's Friend should be studied alongside other physics and philosophy puzzles that feature Restriction~\ref{r:A}, to mutual benefit.
\end{itemize}

Note that there has been previous research on the question of whether certain aspects of WF thought experiments are specifically quantum. Lostaglio and Bowles~\cite{Lostaglio} have shown that the \textit{original} WF thought experiment, involving one Wigner and one Friend, admits a simple classical explanation: Wigner and Friend may simply be two agents with differing knowledge about the same underlying physical configuration. This scenario is widespread and far from mysterious, in particular in classical statistical physics. Furthermore, Hausmann et al.~\cite{Hausmann} have shown that in classical theories such as Spekkens' toy model~\cite{SpekkensToy}, multi-agent paradoxes like Frauchiger and Renner's~\cite{Frauchiger} cannot be reproduced, whereas more general theories such as boxworld, featuring beyond-quantum Bell nonlocality, admit even stronger forms of such paradoxes~\cite{Vilasini}. While these results are important contributions to the WF research program, they do not meet the goal that we are setting ourselves in this paper: they study whether the corresponding theory \textit{admits the statistical prerequisites that are typically used to derive WF phenomenology} (essentially, statistical incompatibilities across different contexts, as in the violation of Bell inequalities), but they do not study directly whether those theories \textit{allow us to draw similar metaphysical or structural conclusions} that the Extended WF thought experiments imply.  Neverthless, our research is closely related to the ``toy model'' program of Spekkens and coauthors, in that we ask which aspects of the EWF arguments can be reproduced classically, versus which cannot -- with the caveat that here we study features of theories, rather than phenomena. We comment more on this in Subsection~\ref{fair_analogy}.\\

\section{Wigner's friends: quantum and classical}
\label{SecWFQC}

In this section, after reviewing the EWF scenario of~\cite{Bong,Wiseman} in Subsection \ref{quantum_experiments}, we describe several classical thought experiments that we argue reproduce main implications of the extended quantum Wigner's Friend experiments (Subsections \ref{classical_experiments} and \ref{reasoning}). In Subsection~\ref{resource_costs}, we argue why these thought experiments ought to be taken seriously alongside their quantum counterparts, and, in Subsection~\ref{fair_analogy}, discuss the relevance of considering these classical analogues in the first place. A thorough structural analysis (in terms of what we call ``Restriction \ref{r:A}'') will follow in the subsequent Section~\ref{SecContext}. For a comparison to existing literature that already relates quantum phenomena with duplication (in particular, Kent's work~\cite{Kent}), see Appendix~\ref{comparison}.

\subsection{The quantum thought experiments}\label{quantum_experiments}

Let us begin by reviewing the thought experiment by Bong et al.~\cite{Bong}.
Their result demonstrates an incompatibility between the controllability of \textit{``quantum evolution [...] on the scale of an observer''} with the conjunction of three assumptions: {Absoluteness of Observed Events}, {Locality} and {No Superdeterminism}:
\begin{enumerate}
    \item \textbf{Absoluteness of Observed Events (AOE)}: An observed event is a real, single event, and not relative to anything or anyone.
    \item \textbf{No Superdeterminism (NSD):} Any set of events on a space-like hypersurface is uncorrelated with any set of freely chosen actions subsequent to that space-like hypersurface.
    \item \textbf{Locality:} The probability of an observable event is unchanged by conditioning on a space-like-separated free choice [...].
\end{enumerate}
Their three-party setup~\cite{Wiseman} concerns two spacelike-separated observers (Alice and Bob) and a friend (Charlie), who is inside a closed laboratory on Alice's wing on the apparatus. Bob and Charlie each hold part of a bipartite system, on which they may make some measurement. First, inside his lab, Charlie makes a measurement, yielding some outcome $c\in\{\pm1\}$. Subsequently, Alice and Bob randomly select one of $N$ inputs, $x\in\{1,...\,,N\}$ and $y\in\{1,...\,,N\}$, determining their measurement settings, which in turn each yield respective outcomes $a\in\{\pm1\}$ and $b\in\{\pm1\}$. These together compose the empirical probability table $\wp(ab|xy)$, for many repeats of the experiment. AOE implies that for every choice of settings $x,y$, there is a probability distribution $P(abc|xy)$ that yields the empirical probabilities as marginal distributions. The protocol dictates that, if Alice selects $x=1$, she will open Charlie's lab and directly ask his measurement outcome, thus setting her own outcome as $a=c$. However, if she selects some $x\neq1$, she will perform a different measurement on Charlie together with the contents of his lab.

The three theory-independent assumptions together lead to following possible empirical probabilities~\cite{Wiseman}:
\begin{equation}\label{LFprobabilities}
\wp(ab|xy)=
\begin{cases}
\sum_{c}\delta_{a,c}P(b|cy)P(c) & \text{if }x=1,\\
\sum_{c}{P}(ab|cxy)P(c) & \text{if }x\neq1,
\end{cases}
\end{equation}
where the only constraints on $P(ab|cxy)$ are Locality and No Superdeterminism (or equivalently, Local Agency~\cite{Wiseman}). Bong et al.\ show that models of the form (\ref{LFprobabilities}) must satisfy various Local Friendliness (LF) inequalities.

\begin{figure*}[t]
\centering 
\includegraphics[trim=0 250 0 0,clip, width=.85\linewidth]{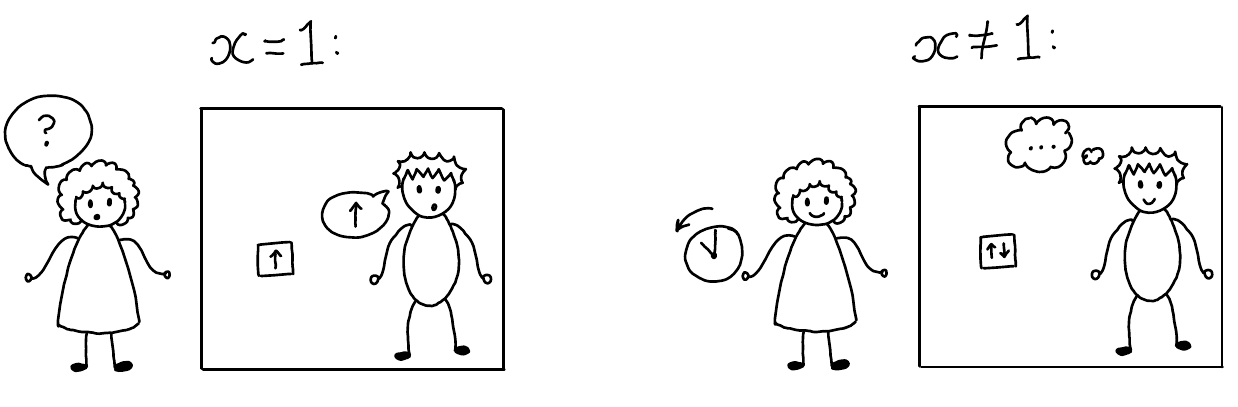}
\caption{Sketch of part of the Local Friendliness setup; Charlie (inside the lab) measures his part of a bipartite system, which is shared with Bob (unpictured). For the setting $x=1$, Alice asks Charlie his outcome. For the setting $x\neq1$, she reverses the contents of the lab, including Charlie, to its ``ready'' state, whereupon she performs a different measurement directly on the subsystem.}
\label{LFfig}
\end{figure*}

The no-go theorem arises from showing that a quantum model exists that violates an LF inequality (part of which is sketched in Figure~\ref{LFfig}). The following situation is considered: Bob and Charlie each hold a qubit, which have been prepared in an entangled state. Charlie performs a measurement on his qubit in a fixed basis, which, according to standard (unitary) quantum mechanics, should be described by a reversible map $U$ acting on the composite Hilbert space of Charlie and his qubit. Next, Alice and Bob choose their input settings $x$ and $y$. Bob accordingly performs one of two measurements on his qubit, which produces outcome $b$. Meanwhile, if $x=1$, Alice asks Charlie his outcome, effectively measuring the qubit in the same basis, such that $a=c$. Or, if $x=2$, Alice reverses Charlie's interaction with the qubit via the inverse map $U^\dag$, and then measures the qubit directly in a different basis, yielding the outcome $a$. It is shown in~\cite{Bong} that there exist states and measurements in quantum theory that lead to probability tables violating LF inequalities -- thus demonstrating a contradiction between quantum theory and the three assumptions. The LF no-go theorem (alongside the many other EWF arguments) may be interpreted as pushing towards perspectival interpretations of quantum theory, in which AOE is rejected -- see, for example,~\cite{Schmid} for a review and analysis of recent EWF arguments, in which the authors note that some form of \textit{absoluteness} is an important ingredient for all EWF arguments. In particular, since EWF arguments all hinge on some formulation of AOE, perspectival interpretations respond to the contradictions in a unified way, by rejecting AOE(-like) assumptions. However, it is certainly possible to reject any of the other assumptions (including the background assumption of quantum-controllability of Charlie), depending on one's favorite interpretation of quantum theory.

The formulation of AOE in~\cite{Bong} takes the notion of an ``observed event'' as a primitive. But when exactly is an event ``observed''? The vagueness of this concept is closely related to the quantum measurement problem, and in physical practice, it is often assumed that we understand what we mean by a ``measurement'' in order to apply quantum theory in the first place. However, once we are interested in studying the consequences of Wigner's Friend beyond quantum theory, it is important to be more specific, and to explain in more detail what ``observed'' is supposed to mean in the prescribed scenarios. This is addressed in a subsequent paper by Wiseman et al.~\cite{Wiseman}, in which they consider the following metaphysical assumptions, with a focus on ``thoughts'': 
\begin{enumerate}
    \item \textbf{Local Agency:} Any [random] intervention [...] is uncorrelated with any set of physical events that are relevant to that phenomenon and outside the future light-cone of that intervention.
    \item \textbf{Physical Supervenience:} Any thought supervenes upon some physical process in the brain (or other information-processing unit as appropriate) which can thus be located within a bounded region in space-time.
    \item \textbf{Ego Absolutism:} My communicable thoughts are absolutely real.
    \item \textbf{Friendliness:} If [...] an independent party displays cognitive ability at least on par with my own, then they have thoughts, and any thought they communicate is as real as any communicable thought of my own.
\end{enumerate} 
\noindent Moreover, they propose the following additional technological assumptions for its implementation:
\begin{enumerate}
    \setcounter{enumi}{4}
    \item \textbf{Human-level AI} can be practically implemented on a digital computer.
    \item \textbf{Universal quantum computing} is physically possible at very large scale, and very fast.
\end{enumerate}
\noindent It is shown in~\cite{Wiseman} that the LF no-go theorem can also be expressed as an incompatibility between quantum theory and the conjunction of the four metaphysical assumptions. In particular, Ego Absolutism states that ``my'' (in the sense of the first person) communicable thoughts are absolutely real -- i.e. my thoughts are objective and need not be qualified relative to anything. Meanwhile, Friendliness states that the communicated thoughts of other intelligent parties are equally as real as my own communicable thoughts. The two together imply that both Wigner's and his Friend's thoughts (which will also contain correlates of their observations) should be taken as absolutely real. In conjunction with Physical Supervenience (that thoughts supervene on physical processes in a bounded region of spacetime), this gives us something metaphysically analogous to AOE. Therefore, when we also assume Local Agency, the contradiction with the predictions of quantum theory (c.f.~\cite{Bong}) can be recovered. Together with the two technological assumptions regarding its implementation, this gives us the ``thoughtful'' LF no-go theorem: one cannot consistently hold all 6 assumptions.

\subsection{Some classical thought experimentation} \label{classical_experiments}

The contradiction presents an important challenge to interpretations of quantum theory, asking which of the six assumptions of~\cite{Wiseman} it is prepared to drop. We would like to make a case though for how similar metaphysical dilemmas arise classically too, simply by considering thought experiments in which persons ``branch''. Our claim is that the notion of ``my'' (in \textit{my communicable thoughts}) can be ambiguous, and that this may be one reason for the failure of the conjunction of the four metaphysical assumptions, quantumly but also classically. That is, one does not need to go to the quantum regime in order to see that the language with which we discuss persons and thoughts is inherently restricted, and runs us into contradictions when taken to more exotic scenarios.

Let us start with a speculative thought experiment. Imagine a world in which humans reproduced via binary fission, c.f.\ the Ebborians~\cite{Yudkowsky,Yudkowsky2}. At some stage in everyone's life, they divide spontaneously into two identical copies of themself, both of whom have psychological and physical continuity~\cite{Parfit} with their prior, singular self. Since the two subsequent persons will go on to be shaped by different experiences, we would naturally conceive of them as two distinct individuals, from the moment of fission. In such a world, we would presumably have developed language to accommodate the fact that a person, who existed singularly in one instance, may now exist as two separate persons. Perhaps, in such a world, we would qualify our references to people spatiotemporally, or perhaps we would simply have a weaker ontological commitment to the notion of persons as persisting entities. In some way though, our language would surely reflect the propensity for persons to branch.

In fact, one of the possible, counterintuitive consequences of quantum theory is that we may, in some sense, already live in such a world. The Everettian response to the measurement problem contends that quantum interactions result in a branching, or duplication, of systems -- including persons. Nevertheless, though our world may genuinely contain branching persons (and on an enormous scale), our emergent, classical view is restricted to only one branch -- so we generally do not run into linguistic problems in referring to our friends who may actually exist in multiplicities. Accordingly, our language has evolved not needing, by and large, to accommodate the possibility for branching persons. As such, we end up hamstrung by semantic oversights, when we consider instances in which branching does occur.

There is already extensive literature in philosophy attempting to give a metaphysical/semantic account of personal identity in branching scenarios~\cite{Parfit,Bishop,Wallace,Lewis,Sider}, as well as real world cases such as split-brain patients~\cite{Parfit2, MacKay} that further motivate such analysis. One of the central challenges is to resolve the apparent contradiction that derives from the transitivity of identity. The problem arises when we ask the following: if a person, let us call her Freya, is duplicated (by binary fission, or via a duplication machine, c.f.\ Parfit~\cite{Parfit}), should we say that she is the ``same person'' as she was prior to duplication? In general, we commit tacitly to the continuity of personal identity (i.e.\ we believe that Freya is the same person as she was 5 years ago), 
\begin{wrapfigure}{r}{0.45\textwidth} 
    \centering
    \includegraphics[trim=210 170 215 140,clip, width=0.6\linewidth]{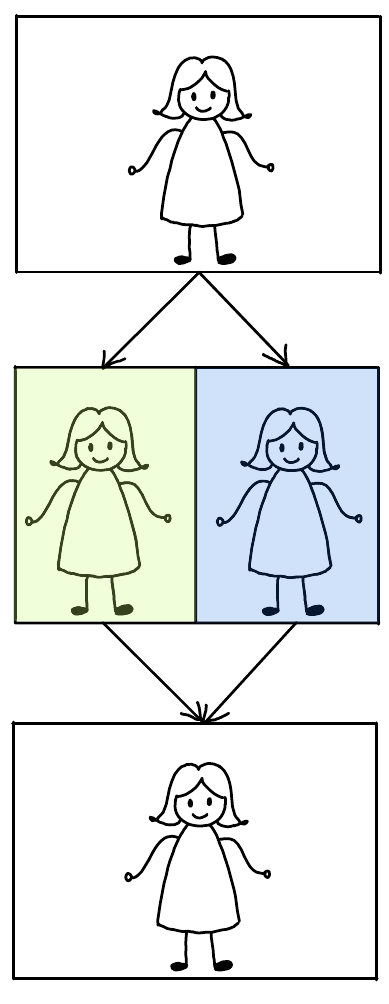}
    \caption{Sketch of Thought Experiment~\ref{te:2}; inside a lab, Freya is duplicated into two identical copies, who wake up in different coloured rooms. This procedure is then reversed, such that a singular Freya wakes up again in a new lab room.}
    \label{TE2fig}
\end{wrapfigure}
which we might cash out more formally in terms of some similarity relation involving physical and psychological continuity. This would have us conclude that {both} Freyas are identical with the previous, singular Freya, since both are physically and psychologically continuous with the Freya that entered the duplication machine. However, accepting that \textit{both} subsequent Freyas are identical with the Freya prior to duplication forces us to resolve that they are also identical with one another, due to the transitivity of the identity relation. But the two Freyas now are causally independent of each other, and will go on to lead different lives -- so such an identification feels mistaken. There are many proposed resolutions to this fallacy (for an overview, please see the Appendix~\ref{branching_identity}, or e.g.~\cite{Bishop} for further details), but ultimately we must accept that our intuitions and language surrounding personal identity are ill-equipped to extend to branching scenarios. 

But why does a philosophical analysis of identity even matter? This analysis certainly does not undermine the derivation for the theory-independent LF no-go theorem~\cite{Bong}, nor the metaphysical choices as presented in~\cite{Wiseman}. However, an articulated account of identity is required in order for us to understand the content of the assumptions that we might choose to discard -- which may moreover be untenable even in a purely classical world. In particular, consider the following:

\vspace{0.1cm}
\textbf{Ego Absolutism:~\cite{Wiseman}} My communicable thoughts are absolutely real.
\vspace{0.1cm}

\noindent Naively understood, there exists an ambiguity regarding what ``{my}'' indexes for branching scenarios. Returning to the example of Freya, who is yet to be duplicated, she reads Ego Absolutism to say that {her} communicable thoughts are absolutely real. This includes her thoughts in that instance, such as ``I am hungry''. It may also be understood to include thoughts she had this morning, such as ``It is raining''. Does it include her future thoughts though? This afternoon, she will be duplicated, whereupon her future copies will have separate experiences. Thus, in describing any future thought she may have, there is an inherent ambiguity as to the meaning of such statements, and whether or not we should take their referent as ``absolute''. That is, it is unclear what the words ``my (future) thoughts'', if uttered by Freya before the start of the experiment, would refer to, and in disregarding this indexical ambiguity, we will typically be led to mathematical formulations of Ego Absolutism that tacitly involve additional assumptions. Note too that the same ambiguity can be found in the \textit{Friendliness} assumption, when ``an independent [and intelligent] party'' is the subject of branching. Not taking this ambiguity into account may easily mislead one to the formal assumption that there is always, at every time, a \textit{single} variable describing a single thought of some person called Freya, while in this branching scenario there are actually two (c.f. the loophole of Kent~\cite{Kent}, see Appendix~\ref{comparison}). This assumption of a unique random variable pertaining to thoughts (which can be described as \textit{events} under Physical Supervenience) is part of the mathematical formulation of Bong et al.~\cite{Bong} -- and the principles of~\cite{Wiseman}, including Ego Absolutism, are constructed to lead to equivalent formal assumptions.

Let us now collect these ideas into the following simple, albeit difficult to implement, classical thought experiment involving both duplication and ``merging'' (c.f.~\cite{Yudkowsky2}):\pagebreak
\refstepcounter{thoughtexperiment}
\begin{mdframed}
\vspace{0.05cm}
\textbf{Thought Experiment \thethoughtexperiment.}\label{te:2} Consider a classical agent, Freya, who prepares to perform an experiment inside a closed laboratory. We refer to Freya at the start of the experiment as Freya$_0$ in order to distinguish her from subsequent versions of herself. In her lab room, Freya$_0$ is duplicated via some classical cloning device, following which one copy (Freya$_B$) wakes up in a blue room, and the other (Freya$_G$) in a green room. Upon awakening, each copy observes which colour room they find themselves in. Next, the two copies are put back to sleep and the memories of their respective experiences erased. They are then merged back together into a singular Freya, who wakes up again in the exit lab room, without any recollection of observations in either colour room in between.
\vspace{0.05cm}
\end{mdframed}

The interventions in the experiment run us into problems concerning personal identity. Freya, reflecting on the experiment, would surely believe herself to be the same person that arrived at the lab to be duplicated, since she has sufficient physical and psychological continuity with Freya$_0$. However, she has no recollection of waking in either blue or green rooms, therefore she cannot identify herself solely with either Freya$_B$ or Freya$_G$. Still, ``forgetting'' is not generally thought to be sufficient to deem someone to be a different person.  Moreover, in all other ways Freya has a very high degree of physical and psychological continuity with the two copies who were in the coloured rooms, and is causally continuous with them. In some sense, she is (/was) \textit{both} Freya$_B$ or Freya$_G$. The truth or falsity of statements such as ``I was in the green room'', ``I was in the blue room'', or even ``I was in the blue \textit{and} the green rooms'' are ambiguous, without clarifying some specific account of personal identity that elucidates a real, singular, persisting referent. On such basis, simply looking at the \textit{wording} of Ego Absolutism and Absoluteness of Observed Events, one may already regard these claims as being violated in the classical thought experiment.

More formally, of crucial importance are the \textit{structural and mathematical} claims that are implicitly associated with the plain language statements. We will examine those in more detail in Section~\ref{SecContext}, but it is instructive to have a preliminary look at those structures and how they relate to the conceptual statements about personal identity made in the previous paragraph. Before Thought Experiment 1 starts, we may imagine that Freya reasons about what she should expect to observe during the experiment. She arguably has reasons to expect the experience of a colourful room; but whether the room will look blue or green (or whether something else, unexpected will be observed) is something that she will be uncertain about. She may want to describe her uncertainty via probability theory: there should be a single random variable that describes the colour $c$ that Freya will see, and a joint probability distribution over all random variables of relevance. For the 4-party setup of Bong et al.~\cite{Bong}, this is explicit in Equation~(\ref{LFprobabilities}) in their paper; AOE is taken to imply mathematically that there is a random variable $c$ (and $d$), which describes the thoughts of the friend(s) throughout the experiment. In their setup, it is ultimately quantum theory that is demonstrated to predict that a random variable of this type and an associated distribution cannot exist, given some natural further assumptions. This is a dramatic conclusion, because it does not only apply to quantum theory as it currently stands, but to all empirically adequate future theories, assuming that the quantum predictions are correct. In our context of Thought Experiment~\ref{te:2}, we make a similar, but more modest observation: classical (statistical) physics, as it currently stands, does not admit a candidate random variable $c$ that could play the desired role (of the ``observed event'') in this experiment, and hence does not predict a probability value that Freya could assign. Following Kent~\cite{Kent}, we might describe it as there being \textit{two} random variables $c_1$ and $c_2$ instead, describing the thoughts of the two Freyas. In this sense, the mathematical formalization of AOE cannot be claimed to hold up in Thought Experiment 1, for \textit{structurally} similar, but \textit{physically} perhaps very different reasons than in Bong et al's quantum scenario.

We do not offer a mathematical proof of this observation; importantly, in contrast to Bong et al., we make no claim that the non-existence of $c$ is true for all other candidate (say, future physical) theories which correctly predict the externally observed outcomes of the experiment, which \textit{would} be a mathematical claim in need of a proof. In our case, our more modest claim follows from the simple observation that saying what Freya will observe after duplication is beyond the range of applicability of classical statistical physics. One way to see this is to acknowledge that all predictions of classical physics are intersubjectively verifiable: all predictions about the movement of material bodies, even if they are probabilistic, can in principle be verified jointly by all agents, i.e.\ from an external perspective. However, in this thought experiment, there is no way to assess ``Freya's thoughts or observations'' from an external perspective. We would not know how to even \textit{define} Freya ``from the outside'' in this thought experiment, which would be needed to define the random variable $c$. In other words, \textit{every} claim of the form ``Freya's private probability of seeing green is $p$'', for \textit{any} $p\in[0,1]$, will be consistent with all predictions of classical mechanics\footnote{Note that we are using the term ``classical mechanics'' rather broadly, allowing also to combine mechanics with probability theory, as in classical statistical mechanics.} as measured by external observers. This brings us back to the conceptual problem of personal identity: defining $c$ would be the formal analogue of defining a notion of personal identity. It would tell us how the first-person concept of ``my'' in the statement of Ego Absolutism is to be translated to a third-person concept, implying how an outside observer would be able to ``measure'' Freya's thoughts or observations during the thought experiment. Classical physics does not tell us how to do that, and hence, does not admit a random variable that would represent Freya's absolute thoughts or observations.

\subsection{When classically duplicated agents reason about each others' reasoning}\label{reasoning}

Another paradigmatic example of EWF arguments is the Frauchiger-Renner Gedankenexperiment~\cite{Frauchiger}, which offers a no-go theorem concerning the consistency of agents' statements when they all reason using quantum theory. The following three (heavily paraphrased) assumptions cannot be jointly valid: 
\begin{itemize}
    \item \textbf{(Q)}: Quantum theory is universally valid;
    \item \textbf{(C)}: The predictions of different agents must be consistent;
    \item \textbf{(S)}: Measurement outcomes must be single-valued.
\end{itemize}

It is found that an agent, upon observing a certain measurement outcome, must simultaneously conclude that another agent has predicted the opposite outcome with certainty. We again argue that some of the metaphysical consequences can be reproduced in a classical example involving duplication of agents, where we will similarly see conflict between different observers' predictions.

To do so, we consider a modified Sleeping Beauty problem, building on a decision-theoretic puzzle that typifies apparent inconsistencies for self-locating beliefs~\cite{Stanford:self-locating}. In its original proposal, an agent is put to sleep and, dependent on the outcome of a coin toss, woken either once or twice; upon each awakening, she is asked for her credence about the outcome of the coin toss, then has her memory erased and is put back to sleep. For a more complete summary, please see Appendix~\ref{sleeping_beauty} and references therein.

\begin{figure*}[t]
\centering 
\includegraphics[angle=270,trim=55 5 200 40,clip, width=1\linewidth]{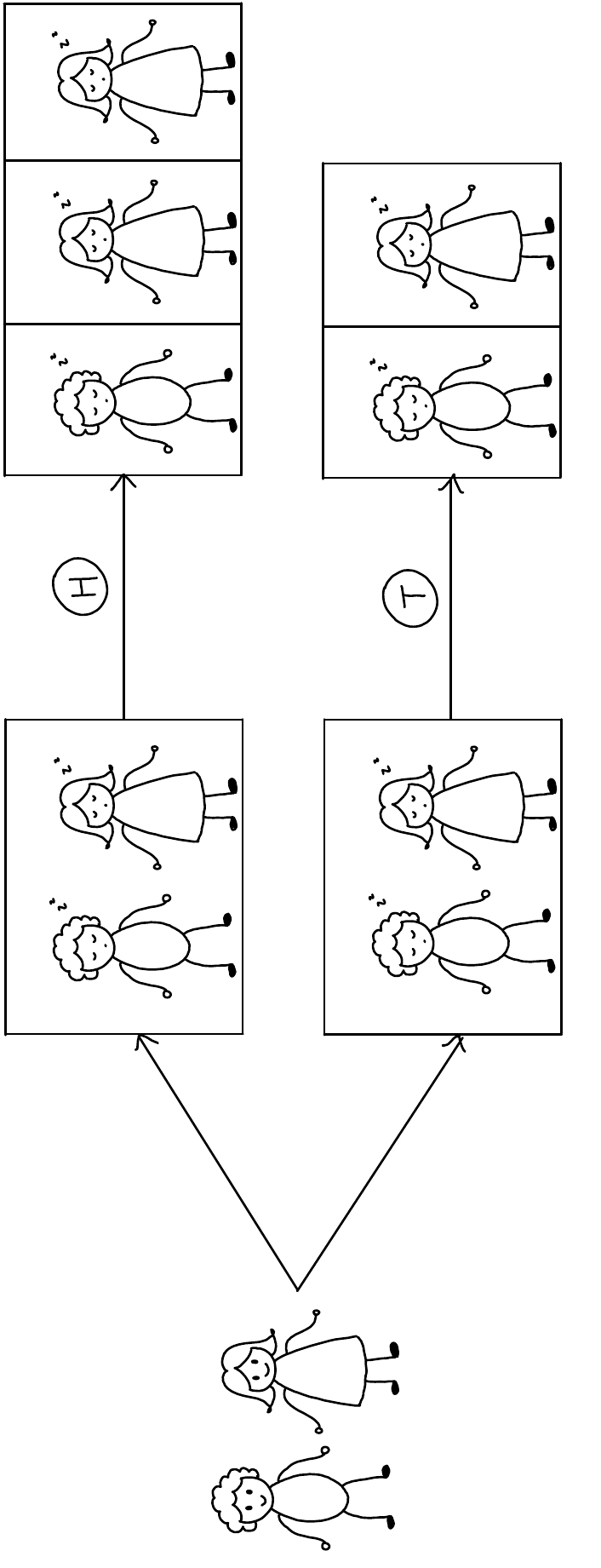}
\caption{Sketch of the setup of Thought Experiment \ref{te:3}; Freya and Wigner agree to an experiment in which they will be put to sleep and multiplied into $N$ copies, where $N$ is large (here though, $N=2$). For each  lab, a fair coin is tossed. If the outcome is Heads, the copy of Freya (but not of Wigner) in the corresponding lab is duplicated again. If the outcome is Tails, she is not.}
\label{TE3fig}
\end{figure*}

\refstepcounter{thoughtexperiment}
\begin{mdframed}
\vspace{0.05cm}
\textbf{Thought Experiment \thethoughtexperiment.}\label{te:3} Imagine Freya and Wigner are to be put to sleep, and multiplied into $N$ copies. Each couple is distributed to one of $N$ identical laboratories. In each laboratory, a fair coin is tossed, and if the outcome is Heads, the copy of Freya (but not the copy of Wigner) is duplicated again. Then, all participants are woken and asked to give their credence that the outcome of their lab's coin toss was Tails. (We assume that they cannot notice the presence/absence of an identical copy of Freya in the lab). This scenario is sketched in Figure~\ref{TE3fig}.

In fact, all participants are offered a bet: they can buy a ticket from a bookie for $(2/3-\varepsilon)$\$, where $\varepsilon>0$ is small, (say, for $66$ cents) that wagers on the coin toss having shown Heads. It is natural to argue (see below) that the credence that Freya should assign to Tails (which directly determines the maximum price $1-p$ she rationally ought to be prepared to pay) is $1/3$, whilst for Wigner it is $1/2$: Freya should buy the ticket, but Wigner should not.

Finally, all copies survive the experiment and are released. Everyone who has bought the ticket now receives $1$\$ if the outcome of their lab's coin toss was indeed Tails. Freya and Wigner have been initially informed about all the details of the experiment.
\vspace{0.05cm}
\end{mdframed}

Given that half as many copies of Freya will experience waking under the outcome Tails than the outcome Heads, a given copy of Freya may assign a proportionately lower degree of belief that she is contained in the smaller group of copies. There is significant disagreement in philosophical literature concerning whether the original Sleeping Beauty ought to be a ``Thirder''~\cite{Elga} or a ``Halfer''~\cite{Lewis2}, and depending on how you operationalise the setup~\cite{Groisman}, one may take either position. Here, we follow the prescription of Elga~\cite{Elga2} for the specific question asked in our Thought Experiment: Elga would claim that Freya should assign a uniform probability distribution to self-locating as any copy (see Appendix~\ref{SubsecIndifference} for an overview on Elga's \textit{Principle of Indifference}~\cite{Elga,Elga2}, and Subsection~\ref{RA_beyond_QT}, in which we reevaluate this option more explicitly).
Using the law of large numbers, there will almost surely be approximately $N$ copies of Freya who will experience a Heads-awakening, following the duplication process, whilst approximately $N/2$ will experience a Tails-awakening (to first order in $N$). Hence, with high probability, approximately $N/2$ copies of Freya will lose their wager of $(2/3-\epsilon)$\$, whilst approximately $N$ copies of Freya will have profited by $(1/3+\epsilon)$\$. The bet is therefore rational for any copy of Freya to accept for any $\varepsilon>0$ (but not for negative $\varepsilon$) -- thus setting her credence for Tails as $1/3$. Similarly, about $N/2$ copies of Wigner will lose their wager, while approximately $N/2$ copies will have profited. Accordingly, Elga's principle implies the following diverging credences for Freya and Wigner respectively:
\begin{equation}
    P_F(T)\approx\frac{N/2}{N+N/2}=\frac{1}{3}, \qquad
    P_W(T)\approx\frac{N/2}{N/2+N/2}=\frac{1}{2}, \label{FreyaWigner_prob}
\end{equation}
where $P_F(T)$ is given by the proportion of ``Tails-Freyas'' (Freyas who will see the outcome Tails) to the total number of Freyas, and likewise for $P_W(T)$.

As in the original Sleeping Beauty problem, the puzzle can be made more dramatic by imagining that Freya will be multiplied into $M$ copies if the outcome of the coin toss is Heads, where $M$ is large (above, we have discussed the case $M=2$). In this case, Freya's credence in the outcome Tails would accordingly be argued to be $1/(M+1)$, which is vanishingly small, whilst Wigner would still surely assign probability $1/2$:
\begin{equation}
    P_F(T)\approx\frac{N/2}{NM/2+N/2}=\frac{1}{M+1}\stackrel{M\to\infty}\longrightarrow 0, \qquad
    P_W(T)\approx\frac{1}{2}. \label{FreyaWigner_prob_largeM}
\end{equation}
As indicated by the ``$\approx$'' signs, the analysis above is not yet fully rigorous: strictly speaking, Elga's principle applies only to situations in which there is a fixed, predetermined, known number of copies of an agent. Here, however, the number of copies depends on the number of Heads results of the $N$ coin tosses. With probability approaching unity for large $N$, this number is about $N/2$ (up to plus or minus $\varepsilon N$ for any small $\varepsilon>0$). Hence, regardless of how we apply Elga's principle to situations with a probabilistically varying number of copies, the above equations should become exact in the limit of $N\to\infty$.

For what follows, however, we will be interested in finite $N$ (in fact, $N=1$). To this end, we need to implement one further assumption:

\textbf{Assumption (NUI)} \textit{(no update on irrelevant information)}. Suppose that Freya knows that she is one of several identical copies in an experiment, and that all copies will undergo the exact same procedure, independently, until the experiment is completed. Then, her predictions for her future observations will not change if she is told which one of the copies she is.

That beliefs should not change in the absence of new evidence is a very intuitive assumption, and it is related to van Fraassen's reflection principle~\cite{vanF}, as remarked by Groisman~\cite{Groisman}. Consider a modification of Thought Experiment~\ref{te:3}, where shortly before the coin tosses, each one of the $N$ Freyas is told the number $j\in\{1,\ldots,N\}$ of the laboratory that they are in. By Assumption (NUI), Freya's probability assignment for large $N$, i.e. $P_F(T)\approx \frac 1 3$, should not change in light of this knowledge. Indeed, consider the copy of Freya in lab $j$ (call her Freya$_j$). From her perspective, what follows is an experiment with a single lab in which, depending on the outcome of a single coin toss, she will be duplicated or not. That is, Freya$_j$ is subject to Thought Experiment~\ref{te:3} for the special case $N=1$, and Assumption (NUI) leads her to keep assigning the probability $P_{F_j}(T)=\frac 1 3$. 

As we see in this example, (NUI) allows us to obtain probabilistic predictions for private future observations under \textit{single} instances of conditional duplication (or multiplication) from the \textit{many-instances} case. Let us show this more generally. Suppose that some random process selects one of $n$ possible outcomes with probabilities $\{p_i\}_{i=1}^n$, and for outcome $i$ Freya will be copied $m_i$ times. Assuming Elga's Principle and (NUI), which probability should Freya assign to experiencing a future in which outcome $i$ was obtained? If Freya is first multiplied into a large number $N$ of copies, and each copy undergoes the process just described, the law of large numbers implies that, with close to unit probability, approximately $N p_i$ of the $N$ outcomes will be $i$, and hence the number $N_i$ of copies of Freya who will observe outcome $i$ is
approximately $m_i N p_i$. The total number of Freyas $N_{\rm tot}$ is approximately $\sum_i m_i p_i N$, and Elga's Principle says that Freya should hence assign probability $N_i/N_{\rm tot}$. This is approximately $c\cdot m_i p_i$, where $c>0$ is a normalization constant such that $\sum_i c m_i p_i=1$. In the limit of $N\to\infty$, this will become exact with unit probability due to the law of large numbers. Now, suppose that Freya is told which of the $N$ copies she actually is, before the random experiment is actually being performed (say, that she is Freya$_j$, where $j\in\{1,\ldots,N\}$). Then (NUI) says that Freya$_j$ should still assign the same probability
\begin{equation}\label{eq:subjective_probs}
    P_F(i)\propto m_i p_i
\end{equation}
in the resulting scenario, which is simply the multiplication experiment described above for only $N=1$ initial copy. In other words: $P_F(i)\propto m_i p_i$ follows even for $N=1$, i.e.\ without the additional $N$-fold multiplication introduced above, if we consider Elga's Principle in conjunction with (NUI).

So far, we have only shown that in the limit of $N\to\infty$, it is true that
\[
    P_F(T)=\frac 1 {M+1},\qquad P_W(T)=\frac 1 2,
\]
and from this we have derived via (NUI) that it must also hold in the special case $N=1$. This is the main preliminary result that we will use for the proof of Theorem~\ref{TheNoGo} below. However, before turning to this, let us do some more checks to see whether Elga's Principle and (NUI) are actually consistent with each other for the experiments described above. In particular, let us see what we can say about all other finite values of $N$, i.e.\ for $2\leq N<\infty$, and check that the result will be consistent with the claimed limit of $N\to\infty$. To do so, we indicate the possible $N$-dependence by $P_F^{(N)}(T)$.

Let us give two different ways of calculating $P_F^{(N)}(T)$; we will see that they give the same result, indicating consistency of Elga's Principle and (NUI). First, we can repeat the argument above and use (NUI) to reduce the fixed-$N$ case to the $N=1$ case, by imagining that every Freya copy is told which of the copies she is before the coins are tossed. As above, this shows that $P_F^{(N)}(T)=P_F^{(1)}(T)=1/(M+1)$.

For a second option of calculating this probability, let us apply Equation~(\ref{eq:subjective_probs}) to the total number of Heads outcomes over all laboratories. In Thought Experiment~\ref{te:3}, for an outcome of $k$ Heads from $N$ labs, there will be $Mk$ copies of Freya in Heads labs, and $N-k$ copies in Tails labs --- therefore there is a total number of $m_k=N+(M-1)k$ copies of Freya, who wake up in the case of $k$ Heads.
Whilst the Binomial probability of this number of Heads is $p_k={N\choose k}/2^N$, the prescription of Equation~(\ref{eq:subjective_probs}) tells us that Freya's probability is proportional to $m_k p_k$ -- i.e. Freya's subjective probability of the outcome $k$ is
\begin{equation}
P_F^{(N)}(k)=c m_k p_k=c\,\frac{N+(M-1)k}{2^{N}}\left(\begin{array}{c} N \\ k\end{array}\right),
\end{equation}
where $c>0$ is the proportionality constant that may depend on $N$. Intuitively (from Elga's perspective), this subjective probability assignment \textit{should} be different from the third-person, ``objective'' probability of $k$ Heads, since the number of Heads dictates how many copies of Freya wake up to experience this outcome. Therefore, by Elga's Principle, larger $k$ outcomes will be weighted proportionately higher by Freya than by an outside party, since they entail more Freya-multiplications.

If $k$ Heads are obtained in total, then we will have $Mk$ Freyas in those Heads labs, among the total number of $m_k$ Freyas. Hence, Elga's Principle dictates that Freya has probability $P_F^{(N)}(H|k)=Mk/(N+(M-1)k)$ of observing Heads. Therefore, Freya's total subjective probability of observing Heads is as follows:
\begin{align}
P_F^{(N)}(H)=\sum_{k=0}^N P_F^{(N)}(k)P_F^{(N)}(H|k) =c\sum_{k=0}^N\frac{N+(M-1)k}{2^{N}}\left(\!\!\begin{array}{c} N \\ k\end{array}\!\!\right)\frac{Mk}{N+(M-1)k}=c\frac{MN}2.
\label{eqBinomial1}
\end{align}
Similarly, we obtain
\begin{align}
P_F^{(N)}(T)=\sum_{k=0}^N P_F^{(N)}(k)P_F^{(N)}(T|k)=c\sum_{k=0}^N\frac{N+(M-1)k}{2^{N}}\left(\!\!\begin{array}{c} N \\ k\end{array}\!\!\right)\frac{N-k}{N+(M-1)k}=c\frac{N}2,
\label{eqBinomial2}
\end{align}
and the constant $c$ is fixed by the requirement $P_F^{(N)}(H)+P_F^{(N)}(T)=1$, resulting in $P_F^{(N)}(H)=M/(M+1)$ and $P_F^{(N)}(T)=1/(M+1)$, \textit{independently of $N$}. That is, Equation~(\ref{FreyaWigner_prob_largeM}) actually holds exactly. This is consistent with our first method of deriving this probability, and demonstrates that Elga's Principle and (NUI) are consistent with each other in the experiment above -- at least for the specific calculations performed (we cannot exclude that there are more complicated scenarios and/or calculations where an inconsistency shows up, but this will be irrelevant for the formulation of our no-go result in Theorem~\ref{TheNoGo} below).

But now, consider the special case of $N=1$. In this case, Freya observes a Heads outcome if and only if Wigner observes a Heads outcome. Nonetheless, assuming Elga's principle and (NUI), Freya and Wigner assign very different probabilities to this single outcome, as we have just shown. An analogy can be drawn to the Frauchiger-Renner Gedankenexperiment. Consider the following probabilistic generalisation of Frauchiger and Renner's assumption (C):

\vspace{0.1cm}
\textbf{Assumption (CP)} \textit{(probabilistic consistency)}. Suppose that agent $A$ has established that \textit{``I am certain that agent $A'$, upon reasoning within the same theory as the one I am using, and having the exact same knowledge of the world as I, is pretty sure that $x=\chi$ at time $t$.''} Then agent $A$ can conclude that \textit{``I am pretty sure that $x=\chi$ at time $t$.''}
\vspace{0.1cm}

If $M$ is large, and if we assume Elga's principle and Assumption (NUI), then Wigner will assign probability $P_W(T)=\frac 1 2$ to the Tails outcome, whereas Freya will assign $P_F(T)=\frac 1 {M+1}$. That is, Freya and Wigner have exactly the same information about the experiment, and they use the exact same theory for prediction. However, whilst Freya is pretty sure that she will observe outcome Heads, Wigner is not. This violates Assumption (CP). While the calculations in equations~(\ref{eqBinomial1}) and~(\ref{eqBinomial2}) show that Elga's principle and (NUI) are \textit{not} in obvious contradiction to each other (at least in Thought Experiment~\ref{te:3}), we have proven incompatibility of both together with Assumption (CP):
\begin{theorem}
\label{TheNoGo}
For Thought Experiment~\ref{te:3}, the following three assumptions cannot jointly hold:
\begin{itemize}
\item Elga's Principle of Indifference;
\item Assumption (NUI) (no update on irrelevant information); 
\item Assumption (CP) (probabilistic consistency).
\end{itemize}
\end{theorem}

This ``classical no-go theorem'' will be analysed more thoroughly in Section~\ref{restrictions}, in particular following Claim~\ref{ClaimFRClassical}.

\subsection{Resource costs of implementation on a classical computer}
\label{resource_costs}

These thought experiments may perhaps be dismissed as science fiction, or at least tangential to scientific inquiry. Whilst EWF arguments push quantum theory to its logical limits in order to explore what may already be happening in reality, our classical thought experiments postulate interventions that are not (so far) realised or realisable. We argue though that they ought to be regarded as equally plausible as some of their quantum counterparts, and moreover that such classical experimentation can similarly be taken to be implementable in the (in fact, much nearer) future. 

Whilst humans are typically envisioned to be the participants in this class of observer puzzles, the authors of the thoughtful LF no-go theorem of~\cite{Wiseman} look at the cost of actually performing their quantum experiments using computer-simulated minds. In the same spirit, we also examine the potential resources and technologies required for duplicating and merging computer-simulated observers -- such that our thought experiments too should not be taken as wild and outlandish, but worthy of being considered in the same context. As an aside, we note that a classical simulation of the mind relies on the assumption that quantum-level information processing is not required for brain emulation -- therefore the following arguments are incompatible with quantum accounts of the mind or consciousness, as proposed by e.g.~\cite{Hameroff,Hameroff2,Chalmers}. Nevertheless, the resource estimates used in~\cite{Wiseman} are based on classical AI algorithms in~\cite{Sandberg,Carlsmith}, which generally follow~\cite{Tegmark} in assuming that the short decoherence timescales preclude quantum phenomena from playing a significant role in brain processing. 

Estimates for the storage costs involved in classically simulating the human brain are typically given anywhere in the range of $10^9$ to $10^{28}$ bits, with~\cite{Wiseman} using $S=10^{15}$ bits as an appropriate space approximation. The required rate is taken to be of the order $F=10^{15}$ FLOP/s (floating point operations per second), or $\gamma=10^{19}$ s$^{-1}$ (gates per unit time). Positing a high degree of parallelisation, this can be taken as a depth per unit time of $\delta=10^{11}$ s$^{-1}$. For the AI to have thoughts at a similar rate to a human (of the order $T=1$), the model is taken to be of depth $t=\delta T=10^{11}$. 

By contrast, implementing such an AI on a quantum computer, the estimates of~\cite{Wiseman} must account for the additional logical and physical quantum resources, as well as fault tolerant quantum gates and error correction. Fault tolerant quantum gates are much slower (7 orders of magnitude) than classical gates. This, combined with the overhead due to gate synthesis (3 orders of magnitude), entails that the time for the AI to have thoughts is of order $T_Q\sim10^{10}$ s (or 500--600 years). Given that the initial target was $T=1$ s, the feasibility of the quantum experiment requires major advances in quantum computing -- possibilities for which the authors discuss in section 6.6 of~\cite{Wiseman}. Meanwhile, the classical implementation of such an AI, with its advantage of at least 7 orders of magnitude, will clearly be feasible far sooner. And so will be the implementation of our classical duplication (or multiplication) thought experiments: processes such as fission, fusion (as in Thought Experiment~\ref{te:2}) or the parallel simulation of $M\approx 10$ copies should slow down the classical AI implementation by a constant factor of not much more than 2 orders of magnitude, which makes it still much faster than the quantum implementation of a single AI.

\subsection{Are these fair analogies?}\label{fair_analogy}

Irrespective of the relative implementability of the quantum and classical experiments, the skeptical reader might be asking, more fundamentally: Why should we be interested in these analogies at all? For most (non-Everettian) quantum theorists, the thought experiments we present here are metaphysically very different from what we may think of as happening during Wigner's Friend scenarios. Why then should we think of them as exhibiting any kind of common features? The skeptic may even accept at this stage that some AOE-like assumption fails to hold in both sets of cases, but maintain that the \textit{core feature}, respectively picked out by the classical and quantum scenarios, is something fundamentally different.

To this point, we wish to highlight the sentiment expressed by Catani et al.~\cite{Catani}, who caution against ``shifting the goalpost'' in reproducing quantum phenomena classically. One should first carefully specify the phenomenon in question (such as the phenomenon of \textit{correlation} or \textit{interference}), and then either prove a no-go theorem that precludes its classical reproduction (as for the possibility of violating a Bell inequality), or provide a scheme to obtain it without quantum theory (as for the phenomena of correlation and interference). There is no point in looking at a classical reproduction and declaring that this is ``not what I actually meant by the phenomenon'', without substantiating precisely what \textit{is} meant. In particular, Catani et al.\ push back on Feynman's famous claim that quantum interference contains the ``essence of quantum theory''~\cite{Feynman}, by showing that there exists a classical toy model that can reproduce the phenomenology of interference experiments without appealing to any of the radical interpretational conclusions.
The authors therefore claim that the traditional explanations of quantum interference (such as wave-particle complementarity, the observer-dependence of reality, and non-local causal influences) should be recognised as deliberate theoretical choices, rather than as being dictated by the appearance of the phenomenon of superposition. (Note that this conclusion applies to \textit{superposition} as a general feature, but not to its \textit{detailed functional form}: for example, the precise trade-off in quantum theory between path distinguishability and fringe visibility \textit{does} constitute a specifically non-classical phenomenon in some precise sense~\cite{Catani2}.) 

It is worth noting the disanalogy between our article and the toy model program of Spekkens~\cite{SpekkensToy}, in that we are not trying to reproduce quantum \textit{phenomena} classically, but rather \textit{features}, or alleged \textit{consequences}, of quantum theory. In this sense, our goal is perhaps more closely related to the research program of e.g.\ Del Santo and Gisin~\cite{DelSanto1,DelSanto2,DelSanto3}, who similarly tackle the question of which features are unique to or characteristic of quantum physics, with their focus on \textit{indeterminism} in classical physics (including a classical analogue of the measurement problem via objective, ontological indeterminacy).

In the same vein, we contend that the supposedly quantum feature captured by Wigner's Friend and the subsequent no-go theorems is not unique to quantum physics. Of course, we do not argue that classical variants of EWF scenarios can reproduce the quantum phenomenology (in particular, by violating an LF inequality): this necessarily involves the violation of a Bell inequality, which is provably incompatible with any classical explanation (in the sense of a local hidden variable model). However, the novelty of the experiment was certainly not the violation of Bell-type inequalities, which is already well established. The novel claim instead is the possibility for violating a combination of metaphysical assumptions -- and the most interesting of those, we argue, are already untenable for analogous classical experiments. Of course, there are many formulations of EWF arguments, constructed from a variety of metaphysical assumptions - some of which may be better suited to identifying some fundamentally nonclassical feature (e.g.\ recently, the authors of~\cite{Walleghem2024} have proposed a GHZ–FR paradox, based on consistency amongst only communicated measurements). Nevertheless, our analysis has shown that there exist undeniable similarities between some EWF scenarios and classical thought experimentation -- and this commonality should not be ignored. To formulate this claim in a more precise way, in Section~\ref{SecContext} we give a definition of the feature that we believe is captured by EWF experiments, and can be classically reproduced. This property, termed ``Restriction \ref{r:A}'', captures the impossibility in some scenarios of obtaining a probability distribution for one or more agents from a given physical theory -- which we see as the common core of Wigner's Friend and other classical puzzles appearing in physics and philosophy.

For Everettians though, our classical thought experiments can be seen as more directly analogous with their quantum counterparts, as Wigner's Friend already contains some kind of duplication of persons. In particular, Wallace~\cite{Wallace} remarks on the philosophical challenges for personal identity in the many-worlds interpretation~\cite{MWI:Stanford}, arguing that branching forces us towards a theory of identity in which \textit{multiple persons} supervene on a single state. This attitude can be discerned in physics literature too, such as~\cite{Deutsch}. Deutsch writes that the formalism of quantum theory is inconsistent with there being an ``actual'' result of a measurement, distinguished from other possible outcomes -- presenting a tension with our experience, which the MWI addresses by positing that observables are typically multivalued, possessing all eigenvalues across a multiplicity of worlds. Applying this proposal to Wigner's Friend entails that the Friend's interaction with a quantum system described by a cat state creates \textit{two copies} of the Friend, each of whom observe one of the two outcomes. An Everettian account of Wigner's Friend is therefore metaphysically very similar to our Thought Experiment \ref{te:2} -- only our persons are side-by-side, rather than displaced across worlds. Furthermore, \cite{Deutsch} proposes an interference experiment involving agents, to test the predictions of many-worlds against collapse interpretations. The observation of interference effects is said to allow the agent to infer that ``there was more than one copy of [themself] (and the atom) in existence'' at the time of the measurement, and that ``these copies merged to form [their] present self''. At this point, the agents are now said to be \textit{once again identical}, despite having previously been in two different branches -- in precise analogy with our Thought Experiment \ref{te:2}. 

This being said, our claim is not ontological in nature -- in particular, we do not try to give an account of what \textit{actually happens} inside the box during a Wigner's Friend experiment. Rather, we claim that the similarities between our thought experiments and certain interpretations of quantum experiments (and, in particular, in their metaphysical consequences) prompt us to ask whether there is a common feature of general physical theories that characterises these puzzles. This leads us to ask about the common, interpretation-independent, structural and mathematical elements of such scenarios, motivating the following section.

\section{Wigner's Friend in context beyond quantum theory}
\label{SecContext}

In this section, we offer a novel way of interpreting the feature at the core of EWF scenarios -- in particular, something structural in nature, that may also apply to other physical theories, e.g.\ classical ones. In Subsection~\ref{restrictions}, we define the concept of \textit{Restriction \ref{r:A}}, show how it applies to the previously described scenarios, and relate it to the existing, metaphysical claims concerning absoluteness~\cite{Bong} or epistemic horizons~\cite{Fankhauser}. In Subsection~\ref{RA_from_RP}, we review how Bell inequality violations in quantum theory can always be lifted to instances of Restriction~\ref{r:A}, and comment on the relation to Fine's Theorem and the relation between agents and physical systems. Finally, in Subsection \ref{RA_beyond_QT}, we discuss the seemingly unrelated Boltzmann brain problem of cosmology in terms of Restriction \ref{r:A}.

\subsection{Definition of Restriction \ref{r:A}}\label{restrictions}

In the previous section, we have argued that important features of Wigner's Friend-type scenarios have analogues in classical duplication experiments. In this section, we will take a more systematic perspective and identify a common structural element of these and other conceptual puzzles, which we define as follows:

\begin{definition}
Suppose we are given a physical theory (for example quantum theory), perhaps together with a set of plausible additional assumptions (such as certain types of locality or causality assumptions). Then ``Restriction \ref{r:A}'' may apply, which we define as follows:

\setcounter{restriction}{0}
\refstepcounter{restriction} 
\textbf{Restriction \therestriction:}\label{r:A} For some experiments, the theory does not provide us with a joint probabilistic description of the observations of all agents involved in the experiment.

Importantly, in this formulation, we assume that the restriction holds even if we are given a complete description of the corresponding experiment within the theory. 
\end{definition}
The term ``joint probabilistic description'' refers to the following formal structure: a probability space $(\Omega,\mathcal{F},P)$, where $\Omega$ is a sample space, a $\sigma$-algebra $\mathcal{F}$ of events, and a probability measure $P$, containing the observations as random variables. Note that Restrictions \ref{r:A} is a possible property of a given \textit{theory} and set of assumptions, rather than of a specific implementation.

Restriction~\ref{r:A} can apply to $N$ agents, and the cases $N=1$ and $N\geq 2$ have very different interpretations: Restriction~\ref{r:A} for $N=1$ means that the respective agent cannot use the theory to obtain a probabilistic prediction for their future observations; Restriction~\ref{r:A} for $N\geq 2$ means that such a prediction may perhaps be obtained, but \textit{not} by predicting the observations of \textit{all} agents and considering the marginal distribution for one of them: the theory does not tell us how the agents' individual observations are correlated, i.e.\ how they fit into a global perspective, formalized by a joint distribution.

Let us first understand a necessary condition for Restriction~\ref{r:A} to apply. To this end, consider an experiment involving several agents, and a theory (say, classical statistical mechanics) that describes it. We can imagine that every agent that is involved in the experiment is initially given a complete description of the experiment. In particular, this will enable the agents to form beliefs about certain facts (such as: what they will see, what the other agents will see, etc.), and such beliefs may be implemented as probability assignments. Accordingly, all agents will start with the same beliefs. However, after some stages of the experiment are concluded, some observers will have learned new facts (obtained some experimental outcomes) that other agents have not. Some of the agents will therefore update their beliefs and some will not -- hence the agents will at some point have different beliefs and statistical descriptions. This is to be expected in essentially all statistical theories. The absence of Restriction \ref{r:A} does \textit{not} imply that all agents will always assign the same probabilities: Restriction \ref{r:A} is something more dramatic than different agents holding different beliefs.

In particular, if Restriction \ref{r:A} applies, then the theory must either fail to describe all statistical empirical observations, \textit{or} it must be impossible for any external observer (say, a physicist who oversees the experiment) to record all observations of all agents and obtain statistics by repeating the experiment. That is, if the latter is possible, then the external observer can obtain a joint statistical description of all agents' observations by experiment, and if Restriction \ref{r:A} applies, then the theory cannot predict these statistics. This would render the theory empirically incomplete. This leads us to the following claim:
\begin{claim}\label{horizon}
    In an empirically complete theory, Restriction~\ref{r:A} only applies in situations where the observations of all agents cannot be externally jointly assessed --- that is, in situations (to borrow the language of \cite{Fankhauser}) that feature an ``epistemic horizon''.
\end{claim}
That is, an essential ingredient for the demonstration of Restriction \ref{r:A} is something that plays the role of a ``black box'' in the experiment. For EWF scenarios, this is the closed laboratory and subsequent unitary reversal of the Friend's measurement process, which forbids the outcome of their measurement being known intersubjectively. Meanwhile, in classical Thought Experiments~\ref{te:2} and~\ref{te:3}, the duplication process makes it impossible to even define from an external point of view what ``Freya's outcome'' has been, even though Freya may privately be justified in wondering before the experiment what she will observe. In this sense, the random experiment becomes a \textit{private experiment}, and the epistemic horizon is due to the fact that its outcome is only known to an agent who performs the experiment themself. Similarly, in the Boltzmann brain (BB) problem (to be discussed further in Subsection \ref{RA_beyond_QT}), the question ``am I a BB?'' may reveal its answer to the questioner, but not to any external observer since the answer does not correspond to any fact of the world.

This also shows that Restriction \ref{r:A} does \textit{not} apply to situations in quantum physics in which all agents are modelled as physical systems and share a common Heisenberg cut: in this case, an external observer can obtain records of all the other agents' observations, since they are on the same side of the cut. Thus, repeating the experiment many times, this observer can empirically estimate the probabilities, and the result must be predicted by quantum theory. This is consistent with Vilasini and Woods' argument that the freedom of choice of Heisenberg cut is an essential element to consider in EWFSs~\cite{VilasiniWoods}. 

As a first example of Restriction \ref{r:A}, consider Bong et al.'s~\cite{Bong} notion of \textbf{Absoluteness of Observed Events (AOE)} (``an observed event is a real single event, and not relative to anything or anyone''): 

\begin{claim}
The negation of AOE is an instance of Restriction \ref{r:A}: the prediction of the violation of a Local Friendliness inequality, as demonstrated in~\cite{Bong}, proves that Restriction \ref{r:A} applies to quantum theory, under the assumptions of Locality\footnote{Note that the Locality assumption of~\cite{Bong} is an assumption only about variables that are \textit{actually observed} (by at least one agent), whereas the locality assumption in Bell's theorem is about \textit{the totality of all (hidden) variables, observed or not}. This also explains why it is insufficient to consider simple scenarios such as two observers (say, Alice and Bob) locally observing the outcomes of a Bell experiment violating the CHSH inequality, to conclude that a joint probabilistic description of their observations via local hidden variables cannot exist, and from this to claim that Restriction \ref{r:A} applies. This conclusion would rest on a less interesting notion of ``Restriction~\ref{r:A}'', defined for quantum theory with a locality assumption about all the hypothetical observations that Alice and Bob \textit{could have made}, rather than on their actual observations.}
and No Superdeterminism. Moreover, it is a particularly strong instance of Restriction \ref{r:A}: every other (say, future) theory that makes the same statistical predictions for the EWF scenario of~\cite{Bong} as quantum theory must also either be subject to Restriction \ref{r:A}, or violate Locality or No Superdeterminism. 
\end{claim}

To see this, assume Locality and No Superdeterminism, and recall the four-party setup of~\cite{Bong} (the three-party version has been described in Subsection~\ref{quantum_experiments} above). The additional assumption of AOE is then equivalent to the claim that for every fixed choice of settings $x,y$, there is a joint probability distribution $P(a,b,c,d|x,y)$~\cite{Bong} on the four observations $a,b,c,d$ made by the four agents Alice, Bob, Charlie, and Debbie, and that Alice's outcome always equals Charlie's outcome if $x=1$, while Bob's outcome always equals Debbie's outcome, if $y=1$. Quantum theory predicts experiments where it at least appears that we can enforce these equality conditions, but observe a violation of a Local Friendliness inequality, implying (under our assumptions) that AOE is false. This implies that there must be at least one pair of settings $x,y$ such that there is no joint probability distribution $P(a,b,c,d|x,y)$ that describes the observations of the four agents. This is an instance of Restriction \ref{r:A}.\qed

That is, the theoretical results of~\cite{Bong} prove that Restriction \ref{r:A} holds for interpretations of quantum mechanics that assume Locality and No Superdeterminism. Moreover, the \textit{experimental} observation of a violation of a Local Friendliness inequality would prove that \textit{all empirically adequate future physical theories} would also be subject to Restriction \ref{r:A} if supplemented by the assumptions of Locality and No Superdeterminism. 

This raises the question of what would count as an experimental demonstration. In fact, the violation of a Local Friendliness inequality has been experimentally demonstrated~\cite{Bong}, but the role of the ``Friend'' was played by the path degree of freedom of a single photon. A single photon is clearly not an agent, but what, then, \textit{is} an agent?

For our purpose, we do not need to give a conclusive answer to the question of what constitutes an agent. Instead, we will work with the following pragmatic and informal definition.
\begin{definition}[Agent]
\label{DefAgent}
When describing a physical scenario (for example, an experiment), we restrict ourselves to using the notion of ``agent'' for an element of the scenario for which any reader (such as You) can in principle consider being this element, at some stage (usually at the initial stage of the experiment). Moreover, we assume that we only describe scenarios where You can expect to have a continued stream of experience (including, for example, observations) over the relevant stages of the experiment. 
\end{definition}
For example, you can imagine being the Friend in a WF experiment and actually putting yourself into the box, before Wigner applies his malicious quantum transformations. While you may have a small degree of apprehension that the drastic microscopic intervention of the WF experiment may render you dead or unconscious, it is not irrational to have high credence in the idea that you will continue your stream of experience (in particular, we can imagine a version of Deutsch's thought experiment~\cite{Deutsch} where the Friend in the box sends the message ``I am alive and well'', instead of ``I see a definite colour''). It is then not irrational to wonder what you will next observe in the box, even though a quantum description of the experiment does not give you an answer to this question, or tells you that even the very existence of an answer in a strict, absolute, naive sense would lead to inconsistencies.

We do not claim that this is a particularly elaborate or insightful definition, but we believe that it is all that is needed for physicists to accept a given thought experiment involving ``agents'' as in principle meaningful.

With this definition in mind, let us consider how Restriction \ref{r:A} relates to Thought Experiment \ref{te:2}:
\begin{claim}\label{claim:te1}
Restriction \ref{r:A} applies to the classical Thought Experiment \ref{te:2}. In more detail, while classical physics may give us a complete description of the thought experiment, and of Wigner's observations, it does not give us a probabilistic description of Freya's observations.
\end{claim}
Should Freya expect to see a blue room or a green room? More specifically, which probabilities should she assign at the beginning of the experiment to these two alternatives? The problem is that it is unclear how to define a random variable (say, $F$) for every stage of the experiment that would describe Freya's observations. From an external perspective (say, by the physicist who constructed the experiment, or by Wigner), there is initially a random variable $F$, and at a later stage, there are perhaps two random variables $F_1$ and $F_2$ describing the properties of the two copies. However, from an internal perspective, it makes perfect sense to be curious to ask ``what will happen to \textit{me} next?''.

We might be inclined to say that the Friend should be indifferent and assign probability $1/2$ to each of the two alternatives. Indeed, Elga has defended a ``Principle of Indifference''~\cite{Elga2}, claiming that one should assign uniform probabilities in cases of self-locating uncertainty. We describe Elga's principle and its relation to our thought experiments in Appendix~\ref{SubsecIndifference}. However, whatever probability value $p$ the Friend decides to put here, this number cannot come from classical physics, but it amounts to an \textit{additional choice} that has to be made. This can also be seen by noting that \textit{it is impossible to verify this probability assignment empirically}: external observers (such as Wigner, and potentially other physicists, in particular those that write publications) cannot repeat the experiment many times and obtain empirical statistics on the number of runs in which Freya ``actually saw'' green versus blue. In particular, we have argued in Section \ref{classical_experiments} that the question itself has no straightforward meaning, since the referent is ambiguous, or even multivalued, if described from an external perspective. The best that Wigner could do is to put himself through the duplication machine and collect statistics that describe his own experience -- but there now exist many different Wigners, each with their own statistical inferences. How do we decide whose statistics should constitute the subjective probability rule for classical physics? (Elga essentially argues that this should be done by averaging over all Wigners, but this comes with its own problem -- see Claim~\ref{ClaimFRClassical}). This is in stark contrast to probability assignments in, say, classical statistical mechanics: all mechanical properties can be measured, statistics can be obtained externally, and thus all probability assignments can in principle be tested empirically\footnote{Here we have assumed that the colour (either green or blue) that Freya will find herself seeing, which we have imagined being described by a probability distribution $(p,1-p)$, is uncorrelated with all facts of the world accessible to external observers. It can be argued that this claim is implicit in the assumption that our scenario involves \textit{duplication}: both resulting versions of Freya (having seen green or blue, respectively) should look like identically legitimate successors of initial Freya for every external observer. For example, both copies should answer ``Freya!'' when asked for their names; similarly, no other facts of the world should have anything to say as to which one of the two is ``more likely initial Freya's successor'' than the other one. In principle, we could imagine that this assumption is not true: perhaps the value of an external variable (say, a classical bit somewhere in thermal radiation) predetermined whether Freya will see green or blue, rendering the other copy's impression that she has had an earlier life as ``Freya before the experiment'' an illusion. But any theory that would make such a prediction is different from, or at least \textit{more than}, classical mechanics, classical statistical mechanics, or any other theory we have previously formulated.}. Hence, the value of $p$ must lie beyond the scope of classical physics. We would need to use ``classical physics \textbf{plus} Elga's Principle of Indifference''~\cite{Elga2}, for example, in order to avoid Restriction \ref{r:A}.

Since aspects of Thought Experiment \ref{te:2} were explicitly constructed as an analogue for Wigner's Friend, it is interesting to compare the two:
\begin{claim}[Non-example: original WF experiment] Restriction \ref{r:A} does \textit{not} apply to the original WF experiment (one Wigner, one Friend). However, if the scenario is extended to allow Wigner to unitarily reverse the Friend's measurement, and to perform a measurement that is incompatible with the Friend's, then Restriction \ref{r:A} applies to these $N=2$ agents. But despite the strong analogy with classical Thought Experiment~\ref{te:2}, Restriction~\ref{r:A} does \textit{not} apply individually to $N=1$ agent (say, the Friend).
\end{claim}
The original WF scenario points out a tension between Wigner's and the Friend's descriptions of the quantum state after the Friend's measurement, but it does not preclude the possibility that the Friend tells Wigner its result after the experiment. Since Wigner and his Friend can (by at least some interpretations) be thought of as sharing a common Heisenberg cut after the Friend's experiment, it is possible to describe all observations made by both by a joint probability distribution, as provided by quantum theory.

However, the situation is different if measurements in incompatible bases are involved (see e.g.~\cite[Section 4.1]{Allam}. It is the possibility of unitarily reversing the Friend's lab (c.f.~\cite{Deutsch}, or in the terminology of~\cite{Schmid}, ``Wigner's Enemy'')  that, under certain assumptions, precludes the existence of a single, consistent probability table -- in particular, because this gives rise to an epistemic horizon~\cite{Fankhauser} (see Claim~\ref{horizon}). The scenario with unitary reversal is analogous to our classical Thought Experiment~\ref{te:2} -- but, in the latter, Restriction~\ref{r:A} even applies to Freya individually: classical physics does not give her a probability for what to see in the experiment, regardless of the question of whether this probability would fit into a joint distribution  with a superobserver. Quantum theory, on the other hand, \textit{does} tell the Friend what to expect to observe before the unitary reversal has been implemented: namely, via the Born rule.

Thought Experiment \ref{te:3} is an even more interesting case of Restriction~\ref{r:A}:
\begin{claim}\label{ClaimFRClassical}
As in Thought Experiment \ref{te:2}, Restriction \ref{r:A} applies to Freya and Wigner individually in Thought Experiment \ref{te:3} (i.e.\ to $N=1$ observer, which may be Freya or Wigner). That is, at the beginning of the experiment, Freya (or Wigner) cannot use classical mechanics for guidance on how she (or he) ought to assign a probability to observing Heads or Tails at the end of the experiment. Moreover, if we supplement classical physics with Elga's Principle of Indifference and (NUI) to inform Freya and Wigner's probabilities, then Restriction A still applies to the $N=2$ observers Freya and Wigner together, i.e. there is no joint probabilistic description for the observations of these two agents, even though each one of them can obtain individual probabilistic predictions from the resulting theory.
\end{claim}
For the same reasons as in Claim~\ref{claim:te1}, classical physics does not provide Freya (and Wigner) with probabilistic predictions for the coin toss results that they will individually observe (recall also our argumentation after Thought Experiment \ref{te:2}). Hence, our scenario contains an instance of Restriction A for $N=1$.

Now suppose we supplement classical physics with a probability rule, such as Elga's Principle of Indifference (or, in fact, any rule for which Wigner's but not Freya's probability is independent of $M$), which defines $P_F(H)$ and $P_W(H)$, and we assume (NUI). As in the proof of Theorem~\ref{TheNoGo}, Assumption (NUI) implies that $P_W(T)$ and $P_F(T)$ are independent of $N$. Hence, consider the case $N=1$. Suppose that Restriction~\ref{r:A} does not hold, then we have a joint probability space with a single distribution $P$ and two events $F_T$ and $W_T$, describing that Freya and Wigner, respectively, observe Tails, reproducing their observation probabilities as $P_F(T)=P(F_T)$ and $P_W(T)=P(W_T)$.
By assumption, since $N=1$, $F_T$ happens if and only if $W_T$ happens, hence $P_W(T)=P(W_T)=P(F_T)=P_F(T)$. But then, if $P_W(T)$ is independent of $M$, so must $P_F(T)$. But, from Elga's principle, this is not the case -- therefore, Restriction~\ref{r:A} must apply.

The application of Elga's Principle of Indifference introduces an observer-dependence into the scenario that can be made explicit by phrasing the question in different words: ``Will I find myself in a world where the coin shows Heads?''. This formulation makes it appear less paradoxial that the transitivity of knowledge that is assumed by (C) and (CP) is no longer relevant; although Wigner may reason ``I am certain that Freya, upon reasoning within the same theory as the one I am using, and having the same exact knowledge of the world as I, is pretty sure that \textit{she will find herself in a world where} $c=H$ at time $t$'', this is not to say that Wigner should conclude ``I am pretty sure that \textit{I will find myself in a world where} $c=H$ at time $t$''. 
This situation resembles others that have been argued to point at an \textit{agent-dependence of facts}~\cite{MWI:Stanford,Dieks,Fuchs,Rovelli,DiBiagio,Healey,Ormrod2}. Moreover, if we interpret Wigner's ``halfer'' and Freya's ``thirder'' assignments as private but objective chances in some sense, then we could regard Freya and Wigner as \textit{probabilistic zombies} relative to each other, as described in~\cite{Mueller}, resembling speculations of~\cite{Sagona-Stophel}.

Whatever resources beyond the physics textbooks Freya consults to obtain a probability assignment that depends on $M$ (such as Elga's Principle of Indifference), this will block her ability to model her observations together with those of other agents (in particular, of Wigner) in terms of a joint probability space, as usually dictated by the postulates of probability theory. In particular, this will block her ability to ``pool'' her knowledge with that of other agents, a phenomenon that has prominently been derived within quantum theory in the Frauchiger-Renner thought experiment~\cite{Frauchiger}. The inability to do so raises further questions, for example a question posed by Renner in the context of quantum theory~\cite{Renner2018}: \textit{Given that Restriction \ref{r:A} blocks several instances of reasoning that combines the knowledge or belief of several agents, what would consitute useful sufficient conditions for when such reasoning is still possible?} For a suggestion of how to address this question in the quantum case, see e.g.\ the work by Renes~\cite{Renes} or Vilasini and Woods~\cite{VilasiniWoods}.

\begin{center}
\textit{Restriction \ref{r:A} from the perspective of personhood}
\vspace{-0.1cm}
\end{center}

In Section~\ref{SecWFQC}, we have argued for the relevance of philosophical positions on personal identity to make sense of the thought experiments. In this subsection, we have so far circumvented this conclusion by relying on a pragmatic definition of ``agent'', see Definition~\ref{DefAgent} above. While this is sufficient to introduce the notion of Restriction \ref{r:A} and discuss its applicability to different physical theories and scenarios, it is still interesting to ask how Restriction \ref{r:A} would be interpreted or understood by proponents of these different philosophical views on personhood, as summarised in Appendix~\ref{branching_identity}.

A possible response by a supporter of Parfit's views could be as follows. They might say that the attempt to regard the ``agent'' in Restriction \ref{r:A} as an ontological existing entity, a \textit{person}, would be misguided and would render this definition meaningless. For example, in Thought Experiment \ref{te:2}, there is no ontological notion of ``Freya'' as the person that entered the experiment \textit{and that also} saw either specific colour. In fact, they might interpret the appearance of Restriction \ref{r:A} as a symptom of the misguided attempt to reason about such a hypothetical entity when there is, as a matter of fact (of the world), actually none, reflected in the structural observation that the world's probability space does not carry a corresponding random variable that persists over the different temporal stages of the experiment. However, they might still agree that Restriction \ref{r:A}, together with our pragmatic definition of agent, is a meaningful notion: Parfit wrote that ``being destroyed and replicated is about as good as ordinary survival''~\cite{Parfit} --- and then, arguably so is the situation involved in a WF-type or duplication experiment. And since it is rational under ordinary survival to attempt to construct theories that help us to assign probabilities to our possible future observations, a Parfitian should then not deny that agents are rational who attempt the same when facing WF-type situations, whatever the word ``our'' means in the \textit{description} of this endeavour.

Sider's `Stage view' might say something similar. Whilst Sider's account of personhood is not as metaphysically reductionist as Parfit's, he too considers the understanding of Freya as a single, persisting person to be mistaken. His own ontology, consisting of three-dimensional person stages, would see there as being multiple persons existing over the course of the experiment. Statements that refer to future or past persons may therefore need to be relativised to a particular class of person stages. Restriction \ref{r:A} may then be viewed as a consequence of mistakenly taking the \textit{aggregates} of person stages to be fundamental (c.f.\ Lewisians, see next paragraph), rather than the stages themselves. In particular, if we consider just the person stage Freya$_0$, prior to the duplication process of Thought Experiment \ref{te:2}, it is perhaps unsurprising that our physical theories do not offer probabilities for whether Freya$_0$ will become Freya$_B$ or Freya$_G$ -- because this question presupposes mistaken notions of persisting persons beyond simple similarity. Equally, for EWF scenarios, perhaps the contradiction is also rooted in an implicit ontology of persons, applied to scenarios for which the identification and composition of person stages is non-trivial. Nevertheless, as with Parfit's view, Freya$_B$ or Freya$_G$ are person-stages that should \textit{matter} to Freya$_0$, due to their connection via an ``{I-relation}'' -- therefore it still seems natural to ask questions about, essentially, \textit{how much} they should matter.

Proponents of the Lewisian `worm view', on the other hand, might say that the type of uncertainty that, for example, Freya faces at the beginning of Thought Experiment \ref{te:2} is nothing but instantaneous self-locating uncertainty: Freya does not know which spacetime worm she is (the one passing through the green room, or the other one passing through the blue room at future spacetime points). At least in the context of classical physics, Lewisians might argue that their view suggests a natural `division of labour': epistemology or the philosophy of mind would be the fields to consult if Freya would like to obtain guidance in the face of her self-locating uncertainty, whereas physics can inform her (and us) about which types of spacetime worms exist, or which ones are probable, according to the laws of (statistical) mechanics. Restriction \ref{r:A} would then be an unavoidable consequence for all theories that we deem physical. However, this conclusion becomes problematic in light of quantum theory: according to some Everettian interpretations, quantum probabilities are of a similar kind to the self-locating uncertainty probabilities (according to Lewisians) in branching scenarios like Thought Experiment \ref{te:2}, and yet, quantum theory as a physical theory has certainly something to say about those probabilities.

In Subsection~\ref{RA_beyond_QT}, we will argue that Restriction \ref{r:A} can be considered an essential ingredient of another puzzle in the foundations of physics and philosophy: the Boltzmann brain problem. Before we turn to this, let us discuss in more detail how Restriction \ref{r:A} is related to Bell nonlocality in quantum theory.

\subsection{Quantum physics: lifting all Bell inequality violations to instances of Restriction \ref{r:A}}\label{RA_from_RP}

Consider a Bell scenario with associated probabilities $P(a,b|x,y)$, where $x,y$ are Alice's and Bob's settings, and $a,b$ are their outcomes. Due to Fine's Theorem~\cite{Fines-theorem} (see also~\cite{Scarani}), if some Bell inequality is violated, then there does not exist a joint probability distribution $P(a_1,\ldots,a_{M_A},b_1,\ldots,b_{M_B})$ which reproduces $P(a,b|x,y)$ via marginalization, where $a_j$ denotes Alice's outcome on setting $j$, and $M_A$, $M_B$ are the numbers of Alice's and Bob's possible settings respectively. Hence, the physical properties $a_j,b_k$ do not have a joint probabilistic description. We can understand the main idea of Bong et al.'s EWFS scenarios as lifting the nonexistence of these joint distributions (which are over \textit{potentially measured} outcomes) to instances of Restrictions A (which is about \textit{actually observed} outcomes).

Violating an LF inequality is a mathematically strictly stronger fact (and hence a metaphysically strictly more dramatic fact) than the violation of a Bell inequality. However, it is instructive to consider the results of~\cite{Utreras-Alarcon}, who have shown that \textit{all} Bell violations can be lifted to some kind of LF violation. In scenarios where we can identify agents with physical subsystems, there is hence a close relation between Bell nonlocality and Restriction \ref{r:A} in quantum theory. However, as we have discussed in the context of Definition~\ref{DefAgent}, and and as is evident in Thought Experiments~\ref{te:2} and~\ref{te:3}, we should not assume that this identification holds in all cases. The result of~\cite{Utreras-Alarcon} is shown by considering \textit{sequential EWFSs}. A sequential EWFS is a scenario in which the superobserving parties (in this case, just Alice) may unitarily reverse the lab and ask their friend to measure in a different basis multiple times prior to ending the experiment. After each measurement performed by Charlie, Alice chooses one of two settings $x_i\in\{0,1\}$, where $1\leq i< R$. If $x_i=0$ and $i<R$, she reverses the lab and asks Charlie to perform a new measurement. Or if $x_i=1$, she asks Charlie his outcome. For the $R$th iteration, after $R-1$ potential reversals, she measures the particle directly for $x_R=0$, or asks Charlie's outcome for $x_R=1$. It is shown that for any two-party scenario $\mathcal{S}=(\mathcal{A},\mathcal{B},\mathcal{X},\mathcal{Y})$, the violation of a Bell inequality within a EWFS implies the violation of a LF inequality, i.e.
\begin{equation} \label{LD=SW}
\mathbb{L}\mathbb{D}(\mathcal{S})= \mathbb{S}\mathbb{W}(\mathcal{S}_{R,0})
\end{equation}
where $\mathbb{L}\mathbb{D}$ denotes the polytope describing behaviours that satisfy local determinism (Bell correlations) and $\mathbb{S}\mathbb{W}$ those that satisfy the LF assumptions for a sequential EWFS, with $R=|\mathcal{X}|-1$. We will now look at the precise definition of these sets.

$\mathbb{L}\mathbb{D}$ is defined as the correlations satisfying \textit{AOE}, \textit{Predetermination}, and \textit{Local Agency}. Within the context of Bell's scenario, these conditions can be expressed as:

\vspace{0.1cm}
{\renewcommand{\arraystretch}{1.3}
\begin{tabular}{r l}
    \textbf{AOE'}:\; & $\exists p(ab|xy),\; \forall x,y$, \\
    \textbf{Predetermination'}:\; & $\exists \lambda, \; p(ab|\lambda xy)\in\{0,1\},\; \forall x,y$, \\
    \textbf{Local Agency'}:\; & $p(a|xy\lambda)=p(a|x\lambda),\; p(b|xy\lambda)=p(b|y\lambda),\; p(\lambda|xy)=p(\lambda),\; \forall a,b,x,y,\lambda$.
\end{tabular}
\vspace{0.1cm}

\noindent This is shown to be equivalent in~\cite{Cavalcanti} to the more traditional formulation of \textit{No Superdeterminism}, \textit{Locality} and \textit{Predetermination} (Bell 1964), or \textit{Local Causality} and \textit{No Superdetermism} (Bell 1976).

$\mathbb{S}\mathbb{W}$ is defined as the correlations satisying \textit{AOE} and \textit{Local Agency}. For a sequential EWFS $\mathcal{S}_{R,0}$, this is expressed as:

\vspace{0.1cm}
\begin{tabular}{r l}
    \textbf{AOE''}:\; & $\exists p(\tilde{a}b\overleftrightarrow{c}|\tilde{x}y),\;\forall     \tilde{x}y$, s.t. \\
    & $p(\tilde{a}|b,\overleftrightarrow{c},\tilde{x}=i,y)=\delta_{a,c_i},\;\forall b,\overleftrightarrow{c},y,1\leq i\leq R$, \\
    \textbf{Local Agency''}:\; & $p(b\overleftarrow{c_j}|x_ix_ky)=p(b\overleftarrow{c_j}|x_ky),\;\forall b,i,y,$\\
    & $k<j\leq i$;\; $p(\tilde{a}\overleftrightarrow{c}|\tilde{x}y)=p(\tilde{a}\overleftrightarrow{c}|\tilde{x}),\;\forall\tilde{a},\overleftrightarrow{c},\tilde{x},y.$
\end{tabular}
\vspace{0.1cm}

\noindent Here, $\tilde{x}$ denotes the first $i$ such that $x_i=1$, determining Alice's final outcome $\tilde{a}:=a_i=c_i$, except in the case where $x_i=0$ for all $i$, for which $\tilde{x}=R+1$. The notation $\overleftrightarrow{c}$ and $\overleftarrow{c_i}$ has also been introduced to denote $\overleftrightarrow{c}=(c_1,...,c_R)$ and $\overleftarrow{c_i}=(c_1,...,c_i)$.

A proof of equation (\ref{LD=SW}) is presented in~\cite{Utreras-Alarcon}.
The first, more established direction is that probabilities in $\mathbb{L}\mathbb{D}(\mathcal{S})$ must also be contained within $\mathbb{S}\mathbb{W}(\mathcal{S}_{R,0})$. This had already been observed in~\cite{Bong}, since the LF set of correlations is a strict superset of the LHV set. 
An intuition for the opposite direction can be obtained by noticing that \textit{AOE''} implies both \textit{AOE'} and \textit{Predetermination'}. 
First, \textit{AOE''} implies \textit{AOE'} for $p(\tilde{a}b|\tilde{x}y)=\sum_{\overleftrightarrow{c}} p(\tilde{a}b\overleftrightarrow{c}|\tilde{x}y)$, when $\tilde{a}$ and $\tilde{x}$ are relabelled $a$ and $x$. 
Second, \textit{AOE''} implies \textit{Predetermination'}, as one can define a hidden variable $\zeta$ containing $\overleftrightarrow{c}$ and $j$, where $j$ labels the extreme points of $\mathbb{N}\mathbb{S}((\mathcal{A},\mathcal{B},R+1,\mathcal{Y}))$.
Charlie's outcomes $\overleftrightarrow{c}$ function as deterministic hidden variables for all but one of the settings chosen by Alice, whilst the distribution over $j$ corresponds to the convex mixture of the different deterministic non-signalling behaviours that are realised in the experiment. Intuitively, it is essentially the claim of the \textit{absoluteness} of Charlie's outcomes $\overleftrightarrow{c}$ that lets them play the role of a set of hidden variables. For details, please see the proof of~\cite{Utreras-Alarcon}.

The equivalence expressed by equation (\ref{LD=SW}) demonstrates that all Bell violations can be lifted to an instance of Restriction \ref{r:A} for quantum theory via an EWF scenario, when one assumes Local Agency. It is also interesting to note that, whilst the quantum thought experiments above all involve instances of violations of Bell-type inequalities, one could also begin with \textit{contextuality}, and lift \textit{this} to examples of Restriction \ref{r:A}. This is done, for example, in Refs.~\cite{Walleghem,Szangolies}.

\subsection{Restriction \ref{r:A} beyond quantum foundations: should you believe you are a Boltzmann brain?}
\label{RA_beyond_QT}

We contend that further puzzles in physics and philosophy concerning identity and first-person experience can be reduced to Restriction \ref{r:A}. In this subsection, we discuss one such example for which we argue that this is the case: the Boltzmann brain (BB) problem. For a brief introduction to this problem, see Appendix~\ref{boltzmann_brain}. Here, we will be even more brief, and discuss only a schematic version ignoring all details that are irrelevant for our discussion. Our exposition will mainly follow Carroll's work~\cite{Carroll}.

Imagine that Freya lives in a universe that is, in the language of cosmologists, ``dominated by Boltzmann brains''. That is, somewhere there is an actual planet Earth containing a human called Freya ($F_0$), but there are \textit{a large number} of copies out there in the universe that are locally indistinguishable from Freya (denote these by $F_1,\ldots,F_N$, where $N$ is large). We assume that these $F_i$ have come into existence by thermal fluctuations: that is, the universe is so large such that we will find enough regions where random processes have led to duplicates of $F_0$ (among many other things that have randomly fluctuated into existence somewhere). Almost all of these $F_i$ will be surrounded by high-temperature thermal fluctuations. Using an illustration from~\cite{Carroll}, if $F_0$-Freya will look at the sky through her telescope in a few minutes from now, she will see the usual microwave background, whereas all the $F_i$-Freyas will see that the microwave background has been replaced by some high-entropy radiation. Note that this observation in itself cannot falsify the possibility for $F_0$-Freya that she is a BB, since a BB-dominated universe predicts that even observers who believe they have made (multiple) corroborating observations are more probably BBs, complete with illusory memories of an imagined, consistent past.

Let us begin by listing the kinds of questions that we will deliberately \textit{ignore} in this paper: how does quantum theory modify our intuition about thermal fluctuations? Should we think of the local reduced state of the universe's vacuum state as ``actually fluctuating'' in some sense, or is it ontologically stationary, given one or another interpretation of quantum theory? Do we have evidence from cosmological observations that supports a picture of the universe that renders it large enough to contain (many) BBs? How should we count the number of BBs at a  fixed time, given that General Relativity does not give us an absolute frame of simultaneity? These questions are best tackled by cosmologists, and some of them are discussed in~\cite{Carroll}.  For our purpose, let us simply argue under the condition that those questions can be considered settled, and ask a different, methodological question: \textit{Is it rational to abandon cosmological models that predict a BB-dominated universe?} That is, are cosmologists correct if they claim that such models are probably false because of one of the following two argumentations:
\begin{itemize}
	\item[(S)] The ``standard argument'': \textit{``[...] in such a universe, I would probably be a Boltzmann Brain, and I'm not, therefore that's not the universe in which we live.''}~\cite{Carroll}
	\item[(C)] Cognitive instability: \textit{``On the one hand, we use our reasoning skills and knowledge of physics to deduce that in such a cosmos we are probably randomly-fluctuated observers, even after conditioning on our local data. On the other hand, we should deduce that we then have no reason to trust those reasoning skills or that knowledge of physics -- thus undermining the basis of our argument}.''~\cite{Carroll}
\end{itemize}
Both (S) and (C) may motivate us to believe that cosmological models which are BB-dominated are false; however, both (S) and (C) rely on the following crucial assumption:\\

\textit{If Freya lives in a BB-dominated universe, she will probably be a BB.}\\

In this paper, we have carefully avoided discussing situations in which agents wonder who or where they \textit{are}, and we have defined the notion of Restriction \ref{r:A} in terms of an agent's (or agents') \textit{observations}. To connect the BB discussion to the one in earlier sections, let us therefore reformulate this statement in a way that makes it more operational:\\

({$*$}) If Freya lives in a BB-dominated universe, then she should expect that she will probably make a BB observation soon} (such as seeing the microwave background being replaced by thermal background).\\

We argue that this reformulation captures more carefully the empirical claim that underlies the standard argument (S) above: after all, we use (what we think are) \textit{our observations} to conclude that we are not BBs. Moreover, quantum physics motivates some caution to ground physical or metaphysical claims in concrete observations as far as possible. For example, the question of whether a photon in an interferometer is in the left or the right arm cannot be given any immediate meaning unless it is operationalised in terms of an empirical observation (say, by asking whether a detector in the left arm does or does not click). Therefore, statement ($*$) is a more careful formulation than the one preceding it.

Statement ($*$) can also be understood as grounding (C), cognitive instability, in a more operational way: if we have randomly fluctuated into existence, then our future observations will probably be uncorrelated with our beliefs and presumed knowledge. In other words: assuming the cosmological model of a BB-dominated universe, ($*$) implies that all our future observations would be uncorrelated with all we know, including this model and its predictions, which in turn should undermine our trust in the model.

However, statement ($*$) is far from obvious. It relies on something like the following unspoken assumption:\\

\textit{If there are $N$ local copies of Freya in the universe, and $M$ of them are BBs, then Freya should expect with a probability of about $M/N$ to make a BB-observation soon.}\\

However, the statement of whether Freya will soon make a BB observation (more succinctly and imprecisely, whether ``she'' is a BB) is not a statement about facts of the world. As such, we cannot use our physical theories directly to assign a probability $p$ to it. More concretely, \textit{physics itself}, i.e.\ our universe's laws and mechanisms, cannot determine its truth value, or the chances of it becoming true. Even if there are laws of the universe that can be phrased in statistical language in the regime of approximation that is relevant for all copies of Freya, these laws will in general fail to imply directly an assignment of probability $p$. The vast probability space describing all relevant facts of the world, i.e.\ whatever classical facts will present themselves to you and your fellow physicists according to your joint choice of Heisenberg cut, will contain $N$ random variables (one for each representation or realisation of Freya), $M$ of them for BBs and $N-M$ of them for non-BBs. But if You, the reader, are ``Freya'', defined by your locally available information and nothing else, then you cannot identify any random variable on this space that would describe \textit{You}. Hence, it is impossible to obtain a marginal distribution from a joint distribution of ``facts of the world'' that would correspond to the correct probability assignment of what You should believe to observe next. This is an instance of Restriction \ref{r:A}. 

We can see the claim above (that one should assign probability $M/N$) as a particularly intuitive attempt to respond to Restriction \ref{r:A} -- that is, to follow Elga's Principle of Indifference~\cite{Elga2} in assigning a uniform probability over all realisations. However, the situation described by Elga (an entertaining scenario involving ``Dr.\ Evil'' summarised in Appendix~\ref{SubsecIndifference}) involves only very few (two) copies of an agent, with the property that the future observations of each copy do not dramatically differ from each other in their information-theoretic properties. It is unclear though whether (or, at the very least, not self-evident that) the uniform distribution is what we should assign in all other cases. In particular, the indifference principle must fail if there are \textit{infinitely many} copies of Freya. Unrealistic as this may seem, we should not exclude this case a priori, since we would like to construct a principled way to respond to Restriction \ref{r:A} that would also work in a literally infinite universe. Furthermore, Everettians in particular may want to admit that (branch or copy) counting is not always what one should do, but other structural features should impact probability assignments~\cite{McQueenVaidman}. Before describing an example of what could potentially replace the indifference principle, let us acknowledge that the above statement ($*$) relies crucially on this assumption, and hence:
\begin{claim}
We \textit{cannot} argue that the prediction of a Boltzmann Brain-dominated universe invalidates a cosmological model, \textit{unless} we respond to Restriction \ref{r:A} in a particular way\footnote{The notion of ``agent'' is here somewhat different from that of the previous thought experiments, but still in accordance with Definition~\ref{DefAgent}. Namely, in our thought experiments and in WF-type experiments, we can at least in its initial stage identify the agent with a fixed, uniquely identified material object. Here, however, the agent is defined as a user of a physical theory that happens to be in a specific local configuration, without specifying which specific piece of matter corresponds to it, or whether (for example) it is itself a Boltzmann brain or not. Still, in line with Definition~\ref{DefAgent}, every reader can in principle imagine putting themselves in the shoes of this element.}.

In other words, we need to make a claim of how to assign probabilities (of agents' future observations) which cannot directly be grounded in our current physical theories, i.e.\ which are not contained in any probability space that would describe the relevant facts of the world, in order to decide whether ``we should believe we are a BB'' in a BB-dominated universe.
\end{claim}
This is the main point we would like to make in this section: Restriction \ref{r:A} is relevant beyond the thought experiment of Wigner's Friend. Its reappearance in other scenarios with direct relevance to physics should be regarded as an argument against an attitude of methodological ignorance or of complete deference to other fields of inquiry such as the Philosophy of Mind.

Finally, let us describe one possible replacement for Elga's Principle of Indifference that has been proposed by one of us in~\cite{Mueller}, which might be called a ``Principle of Induction''. Our goal is not to argue specifically for it, but to point out that there are meaningful options beyond indifference which may lead to different conclusions on the BB problem. Its mathematical formulation relies on the well-known notion of \textit{algorithmic probability}~\cite{LiVitanyi}: if $x$ is any finite binary string (such as $100101$), let $M(x)$ denote the probability that a universal monotone Turing machine with random input will produce a (potentially infinite) output bit string that starts with $x$. Think of Freya's memory being currently described by bit string $x$, and model its memory in the near future by $x$ concatenated with further bits $y$, describing her additional observations. Then, assign a probability proportional to
\begin{equation}
   M(y|x):=M(xy)/M(x)
\end{equation}
to Freya ending up in the copy making future observations described by $y$. For the details of this definition, and the discussion of issues such as the dependence on the choice of universal machine, and lack of normalisation, see~\cite{Mueller}. Note that this is an abstract structural probability assignment, without any ontological claim that there actually exist Turing machines or computations. For the time being, simply note that $M(y|x)$ is large iff $y$ is strongly compressible, given $x$, which can be seen as implementing a Principle of Induction which is indeed applied in what has been termed ``Solomonoff induction''~\cite{LiVitanyi}. For example, if $x=1111\ldots 1$ is a string of $n$ ones, i.e.\ $x=1^n$, then $M(1^m|x)\approx 1$, in the sense that its probability tends to one for $n$ large, for every fixed $m$. On the other hand, if $y_1$ and $y_2$ are both strings of the same length which are not algorithmically related to $x$, then $M(y_1|x)\approx M(y_2|x)$. In this case,  we recover Elga's Principle of Indifference approximately in some special cases. A motivation for the above choice comes from the fact that its predictions are asymptotically consistent with our physical laws in non-exotic, standard laboratory situations (in particular, without duplication) if a physical version of the Church-Turing thesis is true~\cite{Mueller}.

Now, as demonstrated in~\cite{Mueller}, if the above prescription is assumed to define the chances of Freya's future observations, then this particular way to respond to Restriction \ref{r:A} will lead to a consequence that is in some sense the \textit{opposite} to the one drawn via indifference: it implies that Freya should \textit{not} expect to make BB-observations soon, even if she lives in a BB-dominated universe. In a nutshell, the reason for this is that the BB-observation of thermal, uncorrelated radiation makes the resulting data algorithmically uncorrelated with Freya's previous observations, which makes $M(y|x)$ small --- crucially, this value becomes much larger if $y$ describes ``business as usual on Earth''. We refer the interested reader to~\cite{Mueller} for details. This demonstrates the sensitivity of the usual BB argumentation to the choice of ``self-location'' measure, i.e.\ to the way in which we respond to Restriction \ref{r:A}.

\section{Conclusions}
\label{SecConclusions}

In this article, we have presented a property ``Restriction \ref{r:A}'', which we argue is a feature of classical, quantum and indeed more general physical theories: essentially, that a physical theory cannot always provide a probabilistic description of the observations of all agents. We argue that the violation of a Local Friendliness inequality in Extended Wigner's Friend (EWF) scenarios proves a particularly dramatic instance of Restriction \ref{r:A} for quantum theory, and all empirically adequate future theories, if one assumes Locality and No Superdeterminism, but that this should be seen amongst a broader class of puzzles that demonstrate the feature. In particular, we have presented a number of thought experiments involving the duplication of agents, for which classical physics provides no means of assigning outcome probabilities for the users of the theory. Moreover, even supplementing classical physics with some probability rule that supplies these predictions will inevitably lead to inconsistent predictions amongst agents (i.e.\ to a theory that is still subject to Restriction \ref{r:A}), unless this probability rule disregards elements of the setup that are at least intuitively essential. Furthermore, we have shown that Restriction \ref{r:A} is also at the heart of cosmology's Boltzmann brain (BB) problem. In particular, the question of whether cosmological models that predict a BB-dominated universe should be disregarded a priori cannot be answered without reacting to Restriction \ref{r:A} in a particular way.

Moreover, we have shown that the classical duplication experiments reproduce several characteristic features of the quantum WF experiment, and we have related this to the philosophical problem of personal identity. This confirms and extends Kent's analysis~\cite{Kent}, and it may be in line with some supporters of Everettian (many-worlds) interpretations of quantum theory, who might regard the quantum and classical thought experiments as ontologically similar. However, we emphasise that our work does not make any claims about what is ``actually going on in the world'' during a WF-type experiment, and it is not intended to support Everettian views in particular. Our analysis is interpretation-independent, unless stated otherwise. It is also not intended to suggest that extended WF scenarios and their metaphysical conclusions should be unsurprising or irrelevant. Quite the contrary: we suggest that in addition to shedding light on the quantum measurement problem, this research area has important implications that exceed the boundaries of quantum physics.

We take it that the ultimate importance of EWF and similar thought experiments lie in revealing a crucial methodological restriction of our current physical theories: that they typically only provide (probabilistic) predictions for situations in which these predictions can be intersubjectively tested by external observers. However, there are situations such as EWF scenarios for which potential predictions could not be intersubjectively, but \textit{privately} tested: we could certainly imagine putting ourselves into the shoes of the Friend and testing our predictions for what we will observe, but our theories do not always tell us what to expect in situations of this kind. The philosopher Thomas Nagel has famously asked \textit{``What Is It Like to Be a Bat?''~\cite{Nagel}}, emphasizing the significance of the ``subjective character of experience''. In the context of EWF scenarios, we can analogously ask: \textit{What is it like to be a Friend}? In analogy with Nagel's argument, our claim here is that external facts supplied by our physical theories cannot provide a complete picture of internal experiences and predictions. Crucially, this observation does not only refer to the obvious restriction of not knowing \textit{what it feels like} to be a Friend, but even of asking \textit{whether the Friend will hear a detector click or not}. It may be tempting to ignore such questions as ``unphysical'', since their answers cannot be verified by experiments performed by external observers. However, the example of the Boltzmann brain problem demonstrates that this methodological ignorance cannot be sustained indefinitely even within physics: it conflicts with cosmologists' goal to say as much as possible about which models of the universe are plausible.
Moreover, to dismiss questions that we may wish to ask, concerning what \textit{we ourselves} will observe next, as being fundamentally nonsensical seems to deny their genuine relevance for human experience~\cite{Mueller2024}.

Our work suggests several interesting directions for further research. First, we have not said much about the \textit{interpretation} of probability theory that would be sufficient or necessary to provide a conceptual foundation of our argumentation. Myrvold's notion of ``epistemic chance''~\cite{MyrvoldStatMech} may, for example, constitute an adequate underpinning, and it would be interesting to explore this conjecture in detail. Second, our analysis via Restriction~\ref{r:A} challenges Elga's Principle of Indifference (see Theorem~\ref{TheNoGo}), and perhaps (in line with Adlam~\cite{Adlam}) the received view on self-locating credences altogether. This conclusion could be taken as motivation to explore alternative approaches in the foundations of physics that are, broadly speaking, ``idealist'' in nature -- beginning with observations (and their probabilities) as primary notions, and yielding intersubjective events as approximate, emergent features~\cite{Mueller,Mueller2024}.

Besides the question of what to make of Restriction \ref{r:A} in itself, we hope that our work can inspire some novel connections between puzzles that have so far been studied separately in the philosophy and quantum physics communities. We believe that a unified study of this feature can lead to interesting insights that could not be obtained by considering any one of its instances in isolation.

\section*{Acknowledgements} We would like to express our immense gratitude to Laurens Walleghem and an anonymous reviewer, whose very detailed review reports have substantially improved and refined this work. We are indebted to Eric Cavalcanti for numerous detailed explanations of their interesting work~\cite{Bong,Wiseman} by email and discussions in person. We are also grateful to Veronika Baumann, Flavio Del Santo, Andrea Di Biagio, Marius Krumm, Kelvin McQueen, Nuriya Nurgalieva, David Schmid, Howard Wiseman and Y\`il\`e Y\=\i ng for inspiring discussions, as well as helpful comments on the first draft of this manuscript. CLJ would like to thank the participants of the workshop \textit{Kefalonia Foundations 2022: Theoretical and Conceptual Foundations of Quantum Physics} for friendly discussions in a beautiful location. MPM would like to thank the participants of the \textit{Wigner's Friends: Theory Workshop}, Nov.\ and Dec.\ 2022, where some of the ideas of this paper have been presented as a talk, for inspiring discussions, and in particular David Wallace and Adam Brown for helpful discussions on aspects of Everettian quantum mechanics. MPM would also like to thank Mateus Ara\'ujo for literature hints on the many-worlds interpretation. We acknowledge support from the Austrian Science Fund (FWF) via project P 33730-N. This research was supported in part by Perimeter Institute for Theoretical Physics. Research at Perimeter Institute is supported by the Government of Canada through the Department of Innovation, Science, and Economic Development, and by the Province of Ontario through the Ministry of Colleges and Universities.

\section{Appendix}

\subsection{Relation to existing literature}
\label{comparison}

The notion of duplication as a means of interpreting quantum phenomena is already well-established in the literature, in part due to the clear structural similarities with certain interpretations of quantum mechanics.

Notably, and most recently, Kent~\cite{Kent} argues that the LF no-go theorem must be supplemented with an additional assumption that precludes the possibility of duplicated agents. This should be of particular interest to readers of our paper who see the duplication behaviour to be no threat to the assumptions of~\cite{Wiseman}, but see there as being unambiguously two sets of absolute thoughts. Kent comments that variable numbers of agents in the quantum experiment introduce a loophole, through which the derivation of the LF polytope is no longer possible -- therefore a new postulate is needed in order to exclude such interpretations. The ``replicant loophole'' posits that, when he interacts with the subsystem, Charlie is multiplied into two copies, who observe the \textit{pair} of outcomes $c_0$ and $c_1$. These are defined respectively as ``the outcome observed by the Charlie who sees outcome 0'', and ``the outcome observed by the Charlie who sees outcome 1'', such that $P(c_0=0)=P(c_1=1)=1$ by definition. In asking for Charlie's outcome, Alice's measurement destroys the copy of Charlie with whom she did not communicate, and her outcome is set equal to $c_i$, $i\in\{0,1\}$. Whilst her measurement of Charlie previously revealed a pre-existing variable $c$, now Alice's outcome $a$ is determined by $i$, which only comes to exist when she interacts with one of the two Charlies. Accordingly, without some condition $P(a|c,x=1,y)=\delta_{a,c}$ revealing information about some earlier absolute event, the LF probabilities cannot be recovered beyond no-signalling constraints (with which one clearly cannot show any inconsistency with quantum theory). To look at the loophole from the other direction, if one is to observe correlations that violate LF, then the no-go theorem as it currently stands wants to force us to discard (at least) one 
of its assumption -- however, Kent observes that the correlations could also arise from a scenario in which Charlie has been duplicated in his lab, without incurring a contradiction with the original assumptions.

Kent further argues that this bears relevance not just for many-world interpretations, but also arises in single-world versions of quantum theory. For instance, a mass density ontology that posits classically superposed ``half density'' friends (e.g.~\cite{Kent2}) can similarly be seen as exploiting a loophole in the present LF formulation, through which none of the original assumptions need be given up. Charlie can be argued to be split into two consciousnesses up until Alice's measurement, after which there is again just a single consciousness of one ``full density'' friend. Viewing Charlie as briefly existing as two half-density friends entails that, as described above, the LF probabilities cannot obtain the additional structure beyond that of no-signalling probabilities, therefore the LF set is a superset of the quantum set, and no contradiction can be obtained.

To some extent, Kent's insight is a reversal of ours. Kent argues that, when we admit duplication, the LF polytope cannot be recovered, and therefore no inconsistency with quantum physics can be shown. Conversely, we argue that in considering duplication, we already classically violate some of the assumptions of LF (and, by our reading of the theorem, the same assumptions as are undermined in the quantum setting). Therefore, there is not necessarily an inconsistency of quantum physics with the assumptions of LF \textit{any more than there already is classically}. In fact, quantum violations of LF inequalities can be interpreted as violating precisely the assumptions that we have already argued to be untenable for analogous classical situations where friends can be duplicated.

In other literature, thought experiments draw on the types of duplication as is implicit in the Everettian interpretation of quantum physics, in order to describe local and realist (but not based on hidden variables) violations of Bell inequalities. In particular, the \textit{parallel lives} construction of Paul Raymond-Robichaud~\cite{RaymondRobichaud} demonstrates how a local and realist model (not fitting into the hidden variables or ontological models framework~\cite{Harrigan}, and thus not excluded by Bell's theorem) can be consistent with bipartite correlations associated with shared entanglement, or in fact even more general, non-signalling correlations such as those produced by a PR box~\cite{Popescu}. They consider a world in which measurements cause the observers and their surroundings to duplicate into two distinct ``bubbles'', which cannot interact and will never meet again. Two spacelike separated observers perform experiments on respective PR boxes, which cause them to duplicate into bubbles respectively associated with the two possible outcomes. The two \textit{pairs} of observers then meet to compare outcomes, with each bubble keeping only a local memory of which setting was chosen and which outcome observed. When the two pairs of bubbles meet, they interact in a way such that all agents observe statistics consistent with e.g.\ PR boxes (i.e.\ statistics that violate Bell inequalities).

In principle, we could model the full LF experiment via the parallel lives formulation of~\cite{RaymondRobichaud}. In other words, an extreme sort of duplication could be considered such that LF inequalities were violated ``classically'' (in the sense of ``local realism'' as understood by Raymond-Robichaud, but not with a local hidden-variable theory). In the context of their approach, we can interpret these ``classical'' LF violations to similarly undermine the \textit{absoluteness} of facts; outcomes must instead be understood relative to a given ``bubble'' of parallel lives. This leads us to claim, in accordance with the conclusions of the thought experiments considered earlier, that AOE (or Ego Absolutism, for the metaphysical construction) is not tenable even in classical settings if they feature such duplication behaviour. Conversely, others (such as Kent) may view AOE not to be violated by this extended thought experiment, in the sense that there still exists absolute truth values for each agent regarding the outcome they observe and absolute laws governing how agents interact (the caveat being that there is some inherently probabilistic feature of reality that determines which of the two duplicated agents an individual observer will ``become''). By this reading, the LF set of correlations cannot be recovered, as there exists no relevant, single variable $c$ from which to derive the inequality. One would need, as Kent proposes, to supplement the original no-go theorem such to preclude variable numbers of agents. However, by our understanding of AOE, this interpretation does not capture the essence of ``absoluteness'', in the sense that any description, albeit objective, is relativised to a specific bubble. We therefore take a parallel lives construction of the LF scenario to constitute an instance in which AOE is untenable even in a local and realist world.

The theme running through the above literature is that the no-go theorems hinge on the presumption of an event being defined by a \textit{single} hidden variable, and can accordingly be circumvented by discarding this restriction (as in e.g.\ the many-worlds interpretation of quantum theory~\cite{MWI:Stanford}). Moreover, Dieks~\cite{Dieks} offers a realist and local one-world interpretation containing another kind of multiplicity, this time via a \textit{fragmentalist} approach.
These (and other relational, e.g.~\cite{Fuchs,Rovelli,DiBiagio,Healey,Ormrod2}) interpretations give up the assumption of an absolute fundamental description of reality, capturing the perspectives of all observers, as every perspective must be defined {relative to one} of a multiplicity of variables. We have argued that the same \textit{perspectivalism} is sometimes also relevant in a purely classical context, when we consider (sometimes equally reasonable) experiments in which observers can be multiplied. We claim then that we should see these quantum observer puzzles as instances of more general fundamental restrictions on the probabilistic descriptions of agents by physical theories.

\subsection{Accounts of identity for branching scenarios}\label{branching_identity}

Philosophical questions about personal identity evaluate the content of claims such as ``person A is person B'', in particular for instances involving temporal or transworld extent. One of the central challenges for accounts of identity is to resolve the apparent contradiction that arises in branching scenarios from the transitivity of identity, as was outlined briefly in the main text. This bears relevance in thought experiments such as Parfit's teletransportation or many worlds interpretations of quantum mechanics - but also in ordinary, everyday processes, such as an organisms undergoing binary fission or a single zygote dividing into identical twins. We will outline some of the dominant schools of thought for accounts of identity in branching scenarios, although for a more detailed review, we refer the interested reader to~\cite{Parfit,Bishop,Wallace}.

\begin{center}
\textit{Parfit's critique of personal identity}
\vspace{-0.2cm}
\end{center}

Parfit~\cite{Parfit} outlines a series of thought experiments involving teletransportation. First, he considers the simple case in which a Scanner precisely copies all of the data that composes an individual, and reconstructs them on Mars, destroying the original in the process. Science fiction has generally taken this process to be the fastest conceivable form of travel, implicitly assuming that the copy is the \textit{same person} as whoever entered the teletransportation machine. A second thought experiment is then described, in which a modified Scanner does not destroy the original brain and body of the individual on Earth, but just scans them and creates a second, exact copy on Mars. It seems harder here to view teletransportation as a form of transportation, given that there now exist two qualitatively identical individuals, both of whom believe they {are} the same person who entered the machine -- who share all the same memories, intentions and beliefs, and who look and feel the same as they did prior to being scanned. It is asked whether the individual on Earth, upon learning that the Scanner has damaged his cardiac system and that he will die shortly, should be comforted by the fact that his copy on Mars will survive. This involves asking about the nature of identity and what should \textit{matter} to us here regarding survival. 

Parfit presents a reductionist and deflationary account of personal identity. In other words, he {rejects} the views that (1) we are separately existing entities, beyond our brain, body and experiences, and that (2) identity is always determinate, i.e. that there exists real truth-values about statements such as ``person A will be me''. He argues that (2) is indefensible without appealing to (1). Rather, in some cases, questions about identity are \textit{empty} (have no answer). This is highly counterintuitive when it comes to questions about self-identity in branching thought experiments, where, for instance, he argues that there may be no real answer to the question ``am I about to die?''. Parfit argues though that personal identity is not what matters, but a \textit{Relation R}, based on {psychological connectedness and/or continuity}. He further claims that this need not be caused in the normal sense by direct experience, but can have \textit{any} cause (for example, memories that have been generated by a machine). Accordingly, returning to the second thought experiment, the individual survives in the sense that {matters}; there is a surviving copy, with whom he is physically and psychologically continuous.

Perhaps a less convincing aspect of Parfit's account is that identity is generally taken to be determinate, \textit{except} in branching cases with multiple survivors. Bishop~\cite{Bishop} notes that, according to Parfit, it is not the branching itself that kills off individuals, but the by-products; if there is a one-to-one number of replicas, then there is said to be a determinate fact about personal identity, but in the one-to-many case, then no facts of the matter exist. She argues against this position on the basis that ontological crowding should not affect the identity relation. This could be resolved by the (arguably even more radical) view that there is no ontological underpinning for personhood even in the regular, non-branching case, except for physical and psychological continuity. Our memories of the past coupled with our expectation of the future invite us to believe mistakenly that \textit{we} are some further entity that moves through time, but ultimately there is nothing except similarity that unites each time-slice. This adopts a fully antirealist stance on personal identity, and is where Parfit's account ultimately leads us~\cite{Parfit2}: ``Ordinary survival is about as bad as being destroyed
and having a Replica''.

In order to retain a way of talking about personal identity (beyond similarity) over time/worlds, we therefore need to consider other options. Wallace~\cite{Wallace} argues that the two candidates for identity in EQM (and thus, other branching scenarios) are the Worm or the Stage view. He immediately excludes two other possibilities (the Hydra or the Disconnected view), in which personhood itself branches, thus entailing (like Parfit) the rejection of identity altogether. Wallace notes that it is possibly unrealistic to decide between the Worm and Stage views on metaphysical grounds, but they hold semantic differences for EQM.

\begin{center}
\textit{Lewisian `Worm view'}
\vspace{-0.2cm}
\end{center}

In response to Parfit's destructive arguments, the Lewisian account~\cite{Lewis} of personal identity attempts to retain the definiteness and transitivity of personal identity. It is argued that the fallacy arising in scenarios that involve identity over time/worlds stems from the mistaken notion that an individual is wholly located at a given moment. The same issue arises if we take, for example, parts of a person's body to be the whole individual; we now run into similar issues for the identity relation (`the left arm is Freya' and `the right arm is Freya' should not together imply that `the left arm is the right arm', even though reflexivity and transitivity could be used to justify such a claim). However, \textit{parthood} is {not} a transitive relation, and thus the problem can be resolved by reformulating the statements as, for example, `the left arm is part of Freya'. Lewis argues accordingly that a person is a four-dimensional ``worm'' through spacetime, and a three-dimensional time-slice (called a \textit{person-stage}) just composes {part} of the whole individual. The relation of psychological continuity is, by Lewis's terminology, the \textit{I-relation}, which is held between different person-stages of a whole person. This commits Lewis to holding that (in the case of branching) two distinct persons share a common person-stage, prior to branching.

\begin{center}
\textit{Sider's `Stage view'}
\vspace{-0.2cm}
\end{center}

In order to avoid a multiplicity of persons supervening on a singular physical state, Sider's `Stage view'~\cite{Sider} takes just the three-dimensional time-slice to be a person, with no temporal extent. He adopts Lewis's \textit{I-relation} as a means of identifying persons who ``matter'' to you over time; for example, we might say `Freya(2013) is connected to Freya(2023) by a continuous chain of I-related persons'. For branching scenarios, this means that prior to fission there exist multiple future-Freyas that ``matter'' to her. In fact, this stance is analogous to the position adopted by Vaidman~\cite{MWI:Stanford} for the MWI (although he speaks in deflationary terms about identity, sympathising with Parfit's account).

Sider analyses temporal statements in a way analogous with Lewisian counterpart theory of \textit{de re} modality. Accordingly, he argues there is no issue with statements such as `Freya(2023) was 5 years old', even though present-stage 2023 Freya (who did not exist before, and will not exist after) does not have this property. Temporal operators, such as ``was'', are taken to be analogous to the modal operator ``possibly''; they range over temporal worlds in which counterparts of Freya (i.e.\ those that bear the I-relation to present-world Freya) have differing properties, such as being 5 years old. Therefore the truth in the statement `Freya(2023) was 5 years old' lies in the existence of a temporal world in which a counterpart of Freya has the property of being 5 years old. 

In the case of duplication, Sider distinguishes the two statements (a) `Freya$_0$ will be Freya$_B$, and Freya$_0$ will be Freya$_G$' from (b) `Freya$_0$ will be both Freya$_B$ and Freya$_G$'. (b) entails the implausible identity of Freya$_B$ and Freya$_G$, but fortunately the Stage view only implies (a). Therefore the account does not run into problems concerning the transitivity of the identity relation. To see this, we can implement Sider's language of counterpart theory for our Thought Experiment \ref{te:2}. Freya$_0$'s statement ``I will wake up in the blue room'' should be interpreted to mean ``there is a counterpart of me in a future temporal world who will wake up in the blue room'', which is a true statement. Notably, the same statement is also true for the green room (``there is a counterpart of me in a future temporal world who will wake up in the green room''), and therefore the conjunction of the two is also true, c.f.\ statement (a). However, it is not true to say ``there is a counterpart of me in a future temporal world who will wake up in both the blue room and the green room'', c.f.\ statement (b).

\subsection{The Sleeping Beauty problem}\label{sleeping_beauty}

The Sleeping Beauty problem gained significant attention in philosophical literature following a paper by Elga~\cite{Elga} on the decision-theoretic puzzle. The original problem is as follows. Sleeping Beauty (SB) agrees to participate in an experiment, in which she is put to sleep on Sunday, and a fair coin tossed. If the outcome is Heads, she will be woken on Monday only. If the outcome is Tails, she will be woken on Monday, then put back to sleep with her memory erased of Monday's awakening, and woken again on Tuesday. Each of the three possible awakenings are indistinguishable to SB. Upon an awakening, she is questioned about her credence that the coin toss resulted in the outcome Heads.

Opinions on the credence SB ought to assign to the outcome Heads can be broadly divided into two camps; ``Thirders'' and ``Halfers''. The former group (e.g.\ Elga's original paper) argue that the three possible awakenings should be assigned equal probability, according to the \textit{Principle of Indifference}, i.e. $P(H\land \text{Monday})=P(T\land \text{Monday})=P(T\land \text{Tuesday})$. Accordingly, SB should hold a credence in the outcome Heads as $\frac{1}{3}$. The latter group (e.g.~\cite{Lewis2}) argue that SB knew on Sunday that a \textit{fair} coin would be thrown, and being woken has provided her with no new knowledge about the world -- she knew before being put to sleep on Sunday that she would be woken -- therefore her credence in the outcome Heads should still be $\frac{1}{2}$. This is in line with van Fraassen's \textit{Reflection Principle}~\cite{vanF}, that in the absence of new evidence, rational beliefs ought not to change. This is at the core of Bayesian reasoning, yet the SB problem appears to demonstrate an incompleteness of the Bayesian approach when it comes to winning bets~\cite{DutchBook:Stanford}; in order to win under many repeats of the experiment, SB ought to update her beliefs to $\frac{1}{3}$, {despite} the apparent lack of new evidence.

Groisman~\cite{Groisman} deflates the problem by distinguishing two interpretations of the original question, arguing that both Halfer and Thirder arguments are correct, but under two {different setups}. In particular, the credence one ought to assign to the outcome Heads \textit{under the setup of the coin-tossing} is different from that one ought to assign \textit{under the setup of wakening}~\cite{Groisman}. The difficulty in answering the question therefore ultimately comes down to the ambiguity in its phrasing -- whether it refers to SB's credence that the coin lands on Heads or her credence that her awakening is a ``Heads-awakening''. He illustrates this point by means of the following analogy. Imagine again that you toss a fair coin; if the outcome is Heads, you add one green ball into a box, or if the outcome is Tails, you add two red balls to the box. After repeating this procedure many times, you pick a ball at random from the box. Whilst the probability of \textit{adding} a green ball to the box (c.f.\ the coin landing on Heads) is $\frac{1}{2}$, the probability of \textit{selecting} a green ball from the box (c.f.\ a Heads-awakening) tends to $\frac{1}{3}$.

\subsection{Dr.\ Evil and Elga's Principle of Indifference}
\label{SubsecIndifference}

In~\cite{Elga2}, Elga has introduced an entertaining thought experiment to illustrate his principle for self-locating beliefs. It features ``Dr.\ Evil'', residing safely in a battlestation on the Moon, threatening to humanity that he will destroy Earth. Fortunately, Earth is home to a powerful ``Philosophy Defense Force'' (PDF). The PDF sends Dr.\ Evil a message which claims that it has just \textit{``created a duplicate of Dr.\ Evil''}, named Dup, in their skepticism laboratory. The letter continues~\cite{Elga2}: \textit{``At each moment Dup has experiences indistinguishable from those of Dr.\ Evil. For example, at this moment both Dr.\ Evil  and Dup are reading this message. [...] If in the next ten minutes Dup performs actions that correspond to deactivating the battlestation and surrendering, we will treat him well. Otherwise we will torture him.''}

Should Dr.\ Evil surrender? Elga argues in the affirmative: assuming that the PDF is known to have the technological abilities to create duplicates, the reader of this message (be it Dr.\ Evil or Dup) should assign uniform probability of being either Dr.\ Evil or Dup. That is, the reader knows that Dr.\ Evil and Dup are in the same subjective state (his own), therefore he should be unsure as to which of the two copies he is. Accordingly, he ought to assign 50\% probability to being either. Judging his plans not to be worth the 50\% risk of being tortured, Dr.\ Evil should lay down his weapons.

Elga uses this thought experiment to argue for a general ``Principle of Indifference'' in light of self-locating uncertainty. To formulate his principle, he uses the notion of a ``centered world'', which is a possible world with a \textit{``designated individual and time''}~\cite{Elga2}. We can think of this as a possible world together with a choice of ``where (and thus who, or what) I am right now'', for example, world $W$ with Dr.\ Evil on the Moon at time $t$, or world $W$ with Dup on Earth at time $t$. If an agent does not know where they are, \textit{``[they] divide [their] credence among several centered worlds''}.

Elga's Principle of Indifference then reads: \textit{Similar centred worlds deserve equal credence.}

In this postulate, two centred worlds are called ``similar'' if they correspond to the same possible world, and if the designations (the places and times inside this world) are subjectively indistinguishable.

Elga's strategy for defending the Principle of Indifference can be summarised as follows. When we have a duplication process, yielding, say, two identical copies $E$ (Dr.\ Evil) and $D$ (Dup), we can always consider completely unrelated random variables, such as a variable $c\in\{H,T\}$ that describes a coin toss that has nothing to do with the duplication process. We can then think of aggregate events such as $HE,HD,TE,TD$ which say something about the external world (whether the coin toss shows Heads, $H$, or Tails, $T$) and something about self-location (whether ``I'' am $E$ or $D$). Now suppose that our agent learns that some aggregate event $\{HE,TD\}$ has occurred, i.e.\ either $HE$ or $TD$ is the case. After learning this, Elga argues \textit{``[...] his epistemic situation with respect to the coin is just the same as it was before [...]}. \textit{``He has neither gained nor lost information relevant to the toss outcome.''} In other words, the claim is that learning this aggregate event teaches our agent nothing new about the world, and nothing new about the completely unrelated coin toss, and hence $E$ and $D$ must both have the same probability. Otherwise, if, say, $E$ was (much) more likely than $D$, then our agent should believe that the coin shows probably Heads, and his epistemic situation \textit{would} have changed. 

We are agnostic as to whether this argumentation should be considered convincing. It must rely on plausibility assumptions, since the laws of probability theory do not in themselves imply this Principle of Indifference. A perhaps tempting but invalid justification would run as follows: by construction, there is no causal relation whatsoever between the coin toss and whether our agent locates as $E$ or $D$. Thus, learning anything about the coin toss at all, simply by conditioning on the aggregate event $\{HE,TD\}$, is absurd. However, conditioning can make independent events dependent, and this is indeed what happens if $E$ and $D$ have different probabilities. For example, if $E$ and $D$ do not denote self-locations, but ``winning the lottery'' versus ``not winning the lottery'', then conditioning on the event $\{HE,TD\}$ would certainly make our agent learn about the (unrelated) coin toss. Hence, there must be some additional intuition of $E$ and $D$ ``being on equal footing'' entering the argumentation, and this intuition is somehow grounded in the fact that $E$ and $D$ are identical copies.

In our paper, we refer to Elga's Principle of Indifference in some thought experiments that are not literally instances of self-locating uncertainty. For example, in Thought Experiment \ref{te:2}, we ask what initial Freya should believe about whether ``she'' will see a green or a blue room next. Strictly speaking, Elga's principle does not apply: there are never two identical copies who reason at a single time. However, we may imagine that there is a time at which both copies have been created, but have not yet opened their eyes. Elga's principle would then tell us that both should assign probability $50\%$ of being in either room, at this specific time step. If Freya thinks that she will experience being one of the two copies next, then she must hence assign the same probability, since \textit{both} her future copies agree on this. In this sense, we might argue that Elga's principle \textit{can} perhaps be applied. Similar problems of justification of the use of probability theory permeate Everettian interpretations of quantum mechanics~\cite{GreavesMyrvold}.

Here, however, we do not attempt to argue for a final conceptual resolution of this conundrum: in this paper, we are not arguing that the Principle of Indifference (or versions of it) is what one should necessarily apply, but we are only using it as an illustrative example for what one \textit{could} perhaps decide to apply in order to obtain probabilities --- see for example the formulation of Claim~\ref{ClaimFRClassical}.

\subsection{The Boltzmann brain problem}\label{boltzmann_brain}

The notion of Boltzmann brains traces back (surprisingly) to a proposal by Boltzmann -- although the ``problem'' itself was formulated from subsequent arguments. In trying to reconcile the time-reversibility of microphysics with the unidirectional arrow of thermodynamic entropy, Boltzmann proposed that the origin of the observable universe came from a higher entropy state as a \textit{fluctuation}. We therefore happen to find ourselves in an atypical region of a universe otherwise in thermal equilibrium, since low entropy regions uniquely allow for the possibility of conscious observers. However, Eddington~\cite{Eddington} pointed out that the anthropic reasoning present in Boltzmann's argument could equally be applied to smaller fluctuations too -- in particular, whatever is sufficient for an intelligent observer to briefly exist and experience a conscious (coherent) thought. In fact, these fluctuations are exponentially more likely than the type of fluctuations large enough to give rise to the observable universe. Accordingly, he argued that we should expect the universe to be ``in the state of maximum disorganisation which is not inconsistent with the existence of such creatures''~\cite{Eddington}, pushing towards the uncomfortable suggestion that we ourselves are momentary fluctuations from some high entropy state. 

This argument was developed by Albrecht and Sorbo~\cite{AlbrechtSorbo,Albrecht}, who argued that some contemporary models of cosmology predict exponentially more ``Boltzmann brains'' (BBs), minimally consistent with conscious experience and full of memories of an imaginary past, than ordinary observers (OOs), whose thoughts and memories derive from an external universe evolving according to some Hamiltonian. For example, quantum theory predicts that (like black holes) de Sitter spacetime has a horizon, which produces thermal radiation with temperature $T_{dS}=\sqrt{\Lambda/12\pi^2}$. Departing from classical predictions, space then asymptotes to a fixed, nonzero temperature, acting as an eternal thermal system, capable of statistical fluctuations. For exceedingly long timescales, the number of BBs predicted therefore dominates the number of conscious observers in the universe. Accordingly, we ought to assign high credence to being a BB, rather than an OO. 

Not only is this conclusion unappealing, to believe that we have just fluctuated momentarily into existence and will soon dissipate back out, but it also undermines all scientific discourse based on experience -- including the models themselves that predict BBs. BB-dominated models are \textit{cognitively unstable}~\cite{Carroll}, in that they undermine one of their own core assumptions -- that we can trust our sense data and build models accordingly. The argument can be formulated more explicitly as follows. High credence in our sense data and experience (together with many other assumptions) allow us to formulate models of cosmology that describe the lifetime of the observable universe. Certain models (e.g.\ $\Lambda$CDM) predict that there will be many more instances of BBs than OOs in our universe -- therefore it is much more likely that I am a BB than an OO. Being a BB entails that all my senses, experiences and memories are purely illusory, and do not derive from the existence of an external, evolving universe, with a consistent set of physical laws. Therefore, I should not trust my sense data and experience.

A similar line of reasoning is employed by Norton~\cite{Norton}, based on a argument by Myrvold~\cite{MyrvoldStatMech}, in order to show that BBs themselves are self-defeating. In particular, believing that I am a BB undermines all judgements based on my experience -- including statistical mechanics, which forms the basis of the argument for BBs. Nevertheless, whilst this may reassure us that we must be OOs, to avoid logical inconsistency, the threat of cognitive instability remains for models of cosmology that predict too many BBs.

A recent summary of the problem as well as an historical overview can be found in e.g.~\cite{Norton,Carroll}, to which we refer the interested reader for greater depth on the topic.


\begin{thebibliography}{99}

\bibitem{Wigner} E.\ P.\ Wigner, \textit{Remarks on the mind-body question}, In \textit{Philosophical reflections and syntheses}, Springer, 247--260 (1995). \href{https://doi.org/10.1007/978-3-642-78374-6_20}{DOI:10.1007/978-3-642-78374-6\_20}

\bibitem{Maudlin}
T.\ Maudlin, \textit{Three Measurement Problems}, Topoi \textbf{14}, 7--15 (1995). \href{https://doi.org/10.1007/BF00763473}{DOI:10.1007/BF00763473}

\bibitem{BruknerMeasurement}
\v{C}.\ Brukner, \textit{On the quantum measurement problem}, in \textit{Quantum [Un]Speakables II: Half a Century of Bell's Theorem}, Springer (2017). \href{https://doi.org/10.1007/978-3-319-38987-5_5}{DOI:10.1007/978-3-319-38987-5\_5}

\bibitem{Deutsch}
D.\ Deutsch, \textit{Quantum theory as a universal physical theory}, International Journal of Theoretical Physics 24, 1--41 (1985). \href{https://doi.org/10.1007/BF00670071}{DOI:10.1007/BF00670071}

\bibitem{Frauchiger}
D.\ Frauchiger and R.\ Renner, \textit{Quantum theory cannot consistently describe the use of itself}, Nat.\ Commun.\ \textbf{9}, 3711 (2018). \href{https://doi.org/10.1038/s41467-018-05739-8}{DOI:10.1038/s41467-018-05739-8}

\bibitem{Brukner} {\v{C}}.\ Brukner, \textit{A no-go theorem for observer-independent facts}, Entropy,
\textbf{20}(5), 350 (2018). \href{https://doi.org/10.3390/e20050350}{DOI:10.3390/e20050350}

\bibitem{Bong}
K.-W.\ Bong, A.\ Utreras-Alarc\'on, F.\ Ghafari, Y.-C.\ Liang, N.\ Tischler, E.\ G.\ Cavalcanti, G.\ J.\ Pryde, and H.\ M.\ Wiseman, \textit{A strong no-go theorem on the Wigner's friend paradox}, Nat.\ Phys.\ \textbf{16}, 1199--1205 (2020). \href{https://doi.org/10.1038/s41567-020-0990-x}{DOI:10.1038/s41567-020-0990-x}

\bibitem{Kochen}
S.\ Kochen, and E.\ P.\ Specker, \textit{The problem of hidden variables in quantum mechanics}, Ernst Specker Selecta, 235--263 (1990). \href{https://doi.org/10.1007/978-3-0348-9259-9_21}{DOI:10.1007/978-3-0348-9259-9\_21}

\bibitem{Sudbery}
A.\ Sudbery, \textit{Single-world theory of the extended Wigner's friend experiment}, Foundations of Physics, \textbf{47}(5), 658--669 (2017). \href{https://doi.org/10.1007/s10701-017-0082-7}{DOI:10.1007/s10701-017-0082-7}

\bibitem{HealeyWF}
R.\ Healey, \textit{Quantum theory and the limits of objectivity}, Foundations of Physics, \textbf{48}, 1568--1589 (2018). \href{https://doi.org/10.1007/s10701-018-0216-6}{DOI:10.1007/s10701-018-0216-6}

\bibitem{Baumann3}
V.\ Baumann, F.\ Del Santo, and Č.\ Brukner, \textit{Comment on Healey’s ``Quantum theory and the limits of objectivity''}, Foundations of Physics, \textbf{49}, 741--749 (2019). \href{https://doi.org/10.1007/s10701-019-00276-w}{DOI:10.1007/s10701-019-00276-w}

\bibitem{Baumann1}
V.\ Baumann, and Č.\ Brukner, \textit{Wigner’s friend as a rational agent}, Quantum, probability, logic: the work and influence of Itamar Pitowsky, 91--99 (2020). \href{https://doi.org/10.1007/978-3-030-34316-3_4}{DOI:10.1007/978-3-030-34316-3\_4}

\bibitem{DeBrota}
J.\ B.\ DeBrota, C.\ A.\ Fuchs, and R.\ Schack, \textit{Respecting One's Fellow: QBism's Analysis of Wigner's Friend}, Found Phys \textbf{50}, 1859–-1874 (2020). \href{https://doi.org/10.1007/s10701-020-00369-x}{DOI:10.1007/s10701-020-00369-x}

\bibitem{Relano}
A.\ Relaño, \textit{Decoherence framework for Wigner's-friend experiments}, Phys.\ Rev.\ A, \textbf{101}(3), 032107 (2020). \href{https://doi.org/10.1103/PhysRevA.101.032107}{DOI:10.1103/PhysRevA.101.032107}

\bibitem{Guerin}
P.\ A.\ Gu\'erin, V.\ Baumann, F.\ Del Santo, and Č.\ Brukner, \textit{A no-go theorem for the persistent reality of Wigner’s friend’s perception}, Communications Physics, \textbf{4}(1), 93 (2021). \href{https://doi.org/10.1038/s42005-021-00589-1}{DOI:10.1038/s42005-021-00589-1}

\bibitem{Baumann2}
V.\ Baumann, F.\ Del Santo, A.\ R.\ Smith, F.\ Giacomini, E.\ Castro-Ruiz,  and Č.\ Brukner, \textit{Generalized probability rules from a timeless formulation of Wigner's friend scenarios}, Quantum, \textbf{5}, 524 (2021). \href{https://doi.org/10.22331/q-2021-08-16-524}{DOI:10.22331/q-2021-08-16-524}

\bibitem{Leegwater}
G.\ Leegwater, \textit{When Greenberger, Horne and Zeilinger meet Wigner’s friend}, Foundations of Physics, \textbf{52}(4), 68 (2022). \href{https://doi.org/10.1007/s10701-022-00586-6}{DOI:10.1007/s10701-022-00586-6}

\bibitem{Haddara}
M.\ Haddara, and E.\ G.\ Cavalcanti, \textit{A possibilistic no-go theorem on the Wigner's friend paradox}, New J.\ Phys.\ \textbf{25}, 093028 (2023). \href{https://doi.org/10.1088/1367-2630/aceea3}{DOI:10.1088/1367-2630/aceea3}

\bibitem{Walleghem2024}
L.\ Walleghem, R.\ S.\ Barbosa, M.\ Pusey, and S.\ Weigert, \textit{A refined Frauchiger–Renner paradox based on strong contextuality}, Quantum \textbf{10}, 2116 (2026). \href{https://doi.org/10.22331/q-2026-05-26-2116}{DOI:10.22331/q-2026-05-26-2116}

\bibitem{Xu}
Z.\ P.\ Xu, J.\ Steinberg, H.\ C.\ Nguyen, and O.\ Gühne, \textit{No-go theorem based on incomplete information of Wigner about his friend}, Phys.\ Rev.\ A, \textbf{107}(2), 022424 (2023). \href{https://doi.org/10.1103/PhysRevA.107.022424}{DOI:10.1103/PhysRevA.107.022424}

\bibitem{Lostaglio}
M.\ Lostaglio and J.\ Bowles, \textit{The original Wigner's friend paradox within a realist toy model}, Proc.\ R.\ Soc.\ A.\ \textbf{477}, 20210273 (2021). \href{https://doi.org/10.1098/rspa.2021.0273}{DOI:10.1098/rspa.2021.0273}

\bibitem{Hausmann}
L.\ Hausmann, N.\ Nurgalieva, and L.\ del Rio, \textit{Toys can't play: physical agents in Spekkens' theory}, New J.\ Phys.\ \textbf{25}, 023018 (2023). \href{https://doi.org/10.1088/1367-2630/acb3ef}{DOI:10.1088/1367-2630/acb3ef}

\bibitem{SpekkensToy}
R.\ W.\ Spekkens, \textit{Evidence for the epistemic view of quantum states: A toy theory}, Phys.\ Rev.\ A \textbf{75}, 032110 (2007). \href{https://doi.org/10.1103/PhysRevA.75.032110}{DOI:10.1103/PhysRevA.75.032110}

\bibitem{Vilasini}
V.\ Vilasini, N.\ Nurgalieva, and L.\ del Rio, \textit{Multi-agent paradoxes beyond quantum theory}, New J.\ Phys.\ \textbf{21}, 113028 (2019). \href{https://doi.org/10.1088/1367-2630/ab4fc4}{DOI:10.1088/1367-2630/ab4fc4}

\bibitem{Wiseman}
H.\ M.\ Wiseman, E.\ G.\ Cavalcanti, and E.\ G.\ Rieffel, \textit{A ``thoughtful'' Local Friendliness no-go theorem: a prospective experiment with new assumptions to suit}, Quantum \textbf{7}, 1112 (2023). \href{https://doi.org/10.22331/q-2023-09-14-1112}{DOI:10.22331/q-2023-09-14-1112}

\bibitem{Kent}
A.\ Kent, \textit{Friendly thoughts on thoughtful friendliness}, arXiv:2302.12707 (2023). \href{https://doi.org/10.48550/arXiv.2302.12707}{DOI:10.48550/arXiv.2302.12707}

\bibitem{Schmid}
D.\ Schmid, Y.\ Y\=\i ng, and M.\ Leifer, \textit{A review and analysis of six extended Wigner's friend arguments}, arXiv:2308.16220 (2023). \href{https://doi.org/10.48550/arXiv.2308.16220}{DOI:10.48550/arXiv.2308.16220}

\bibitem{Yudkowsky}
E.\ Yudkowsky, \textit{Where Physics Meets Experience}, LessWrong (2008). \url{https://www.lesswrong.com/posts/WajiC3YWeJutyAXTn/where-physics-meets-experience}

\bibitem{Yudkowsky2}
E.\ Yudkowsky, \textit{The Anthropic Trilemma}, LessWrong (2009). \url{https://www.lesswrong.com/posts/y7jZ9BLEeuNTzgAE5/the-anthropic-trilemma}

\bibitem{Parfit}
D.\ Parfit, \textit{Reasons and persons}, OUP Oxford (1984). \href{https://doi.org/10.1093/019824908X.001.0001}{DOI:10.1093/019824908X.001.0001}

\bibitem{Bishop} 
S.\ A.\ M.\ Bishop, \textit{Identity and Counterparthood in a Many Worlds Universe}, PhD thesis, City University of New York (2020). \url{https://academicworks.cuny.edu/gc_etds/3575}

\bibitem{Lewis} 
D.\ K.\ Lewis, \textit{Survival and identity}, In Amelie Oksenberg Rorty, editor, \textit{The Identities of Persons}, 17--40. University of California Press (1976). \href{https://doi.org/10.1525/9780520353060-002}{DOI:10.1525/9780520353060-002}

\bibitem{Sider} 
T.\ Sider, \textit{All the world’s a stage}, Australasian Journal of Philosophy, \textbf{74}(3):433–453 (1996). \href{https://doi.org/10.1080/00048409612347421}{DOI:10.1080/00048409612347421}

\bibitem{Wallace} 
D.\ Wallace, \textit{The emergent multiverse: Quantum theory according to the Everett interpretation}, Oxford University Press (2012). \href{https://doi.org/10.1093/acprof:oso/9780199546961.001.0001}{DOI:10.1093/acprof:oso/9780199546961.001.0001}

\bibitem{Parfit2}
D.\ Parfit, \textit{Divided minds and the nature of persons}, Science Fiction and Philosophy: From Time Travel to Superintelligence, 91--98 (2016). \href{https://doi.org/10.1002/9781118922590.ch8}{DOI:10.1002/9781118922590.ch8}

\bibitem{MacKay}
D.\ M.\ MacKay, and V.\ MacKay, \textit{Explicit dialogue between left and right half-systems of split brains}, Nature \textbf{295}(5851), 690--691 (1982). \href{https://doi.org/10.1038/295690a0}{DOI:10.1038/295690a0}

\bibitem{Stanford:self-locating}
A.\ Egan, and M.\ G.\ Titelbaum, \textit{Self-Locating Beliefs}, The Stanford Encyclopedia of Philosophy (Winter 2022 Edition), Edward N.\ Zalta \& Uri Nodelman (eds.). \url{https://plato.stanford.edu/archives/win2022/entries/self-locating-beliefs/}

\bibitem{Elga}
A.\ Elga, \textit{Self-locating belief and the Sleeping Beauty problem}, Analysis \textbf{60}(2), 143--147 (2000). \href{https://doi.org/10.1093/analys/60.2.143}{DOI:10.1093/analys/60.2.143}

\bibitem{Lewis2}
D.\ Lewis, \textit{Sleeping beauty: reply to Elga}, Analysis \textbf{61}(3), 171--176 (2001). \href{https://doi.org/10.1093/analys/61.3.171}{DOI:10.1093/analys/61.3.171}

\bibitem{Groisman}
B.\ Groisman, \textit{The end of Sleeping Beauty's nightmare}, The British Journal for the Philosophy of Science \textbf{59}(3), 409--416 (2008). \href{https://doi.org/10.1093/bjps/axn015}{DOI:10.1093/bjps/axn015}

\bibitem{Elga2}
A.\ Elga, \textit{Defeating Dr.\ Evil with self-locating belief}, Philos.\ Phenomenol.\ Res.\ \textbf{69}(2), 383--396 (2004). \href{https://doi.org/10.1111/j.1933-1592.2004.tb00400.x}{DOI:10.1111/j.1933-1592.2004.tb00400.x}

\bibitem{vanF}
B.\ van Fraassen, \textit{Belief and the Will}, The Journal of Philosophy \textbf{81}(5), 235--256 (1984). \href{https://doi.org/10.2307/2026388}{DOI:10.2307/2026388}

\bibitem{Hameroff}
S.\ Hameroff, and R.\ Penrose, \textit{Orchestrated reduction of quantum coherence in brain microtubules: A model for consciousness}, Mathematics and computers in simulation \textbf{40}(3--4), 453-480 (1996). \href{https://doi.org/10.1016/0378-4754(96)80476-9}{DOI:10.1016/0378-4754(96)80476-9}

\bibitem{Hameroff2}
S.\ Hameroff, and R.\ Penrose, \textit{Consciousness in the universe: A review of the ‘Orch OR’theory}, Physics of life reviews \textbf{11}(1), 39--78 (2014). \href{https://doi.org/10.1016/j.plrev.2013.08.002}{DOI:10.1016/j.plrev.2013.08.002}

\bibitem{Chalmers}
D.\ J.\ Chalmers, and K.\ J.\ McQueen, \textit{Consciousness and the collapse of the wave function}, in \textit{Consciousness and Quantum Mechanics}, Shan Gao (ed.), Oxford University Press, Oxford, 2022. \href{https://doi.org/10.1093/oso/9780197501665.003.0002}{DOI:10.1093/oso/9780197501665.003.0002}

\bibitem{Carlsmith}
J.\ Carlsmith, \textit{How Much Computational Power Does It Take to Match the Human Brain?}, Open Philanthropy Project Research Memo (2020). \url{https://www.openphilanthropy.org/brain-computation-report}

\bibitem{Sandberg}
A.\ Sandberg and N.\ Bostrom, \textit{Whole brain emulation: a roadmap}, Future of Humanity Institute, Oxford University, Technical Report 2008-3 (2008).

\bibitem{Tegmark}
M.\ Tegmark, \textit{Importance of quantum decoherence in brain processes}, Physical review E \textbf{61}(4), 4194 (2000). \href{https://doi.org/10.1103/PhysRevE.61.4194}{DOI:10.1103/PhysRevE.61.4194}

\bibitem{Catani}
L.\ Catani, M.\ Leifer, D.\ Schmid, and R.\ W.\ Spekkens, \textit{Why interference phenomena do not capture the essence of quantum theory}, Quantum \textbf{7}, 1119 (2023). \href{https://doi.org/10.22331/q-2023-09-25-1119}{DOI:10.22331/q-2023-09-25-1119}

\bibitem{Feynman}
R.\ P.\ Feynman, R.\ B.\ Leighton, and M.\ L.\ Sands, \textit{The Feynman Lectures on Physics} Addison-Wesley world student series, (1961--1963).

\bibitem{Catani2}
L.\ Catani, M.\ Leifer, G.\ Scala, D.\ Schmid, and R.\ W.\ Spekkens, \textit{Aspects of the phenomenology of interference that are genuinely nonclassical}, Phys.\ Rev.\ A \textbf{108}, 022207 (2023). \href{https://doi.org/10.1103/PhysRevA.108.022207}{DOI:10.1103/PhysRevA.108.022207}

\bibitem{DelSanto1}
F.\ Del Santo, and N.\ Gisin, \textit{Physics without determinism: Alternative interpretations of classical physics}, Physical Review A, \textbf{100}(6), 062107 (2019). \href{https://doi.org/10.1103/PhysRevA.100.062107}{DOI:10.1103/PhysRevA.100.062107}

\bibitem{DelSanto2}
F.\ Del Santo, \textit{Indeterminism, causality and information: Has physics ever been deterministic?}, Undecidability, Uncomputability, and Unpredictability, 63--79 (2021). \href{https://doi.org/10.1007/978-3-030-70354-7_5}{DOI:10.1007/978-3-030-70354-7\_5}

\bibitem{DelSanto3}
F.\ Del Santo, and N.\ Gisin, \textit{Potentiality realism: A realistic and indeterministic physics based on propensities}, Eur.\ J.\ Philos.\ Sci.\ \textbf{13}, 58 (2023). \href{https://doi.org/10.1007/s13194-023-00561-6}{DOI:10.1007/s13194-023-00561-6}

\bibitem{MWI:Stanford}
L.\ Vaidman, \textit{Many-Worlds Interpretation of Quantum Mechanics}, The Stanford Encyclopedia of Philosophy (Fall 2021 Edition), Edward N.\ Zalta (ed.). \url{https://plato.stanford.edu/archives/fall2021/entries/qm-manyworlds/}

\bibitem{Fankhauser}
J.\ Fankhauser, T.\ Gonda, and G.\ D.\ L.\ Coves, \textit{Epistemic Horizons From Deterministic Laws: Lessons From a Nomic Toy Theory}, Synthese \textbf{205}, 136 (2025). \href{https://doi.org/10.1007/s11229-024-04852-0}{DOI:10.1007/s11229-024-04852-0}

\bibitem{VilasiniWoods}
V.\ Vilasini and M.\ P.\ Woods, \textit{A general quantum circuit framework for Extended Wigner's Friend Scenarios: logically and causally consistent reasoning without absolute measurement events}, arXiv:2209.09281 (2024). \href{https://doi.org/10.48550/arXiv.2209.09281}{DOI:10.48550/arXiv.2209.09281}

\bibitem{Allam}
J.\ Allam and A.\ Matzkin, \textit{Are Unitary Accounts of Quantum Measurements in Relativistic Wigner’s Friend Setups Compatible in Different Reference Frames?}, Metrology \textbf{4}(3), 364--373 (2024). \href{https://doi.org/10.3390/metrology4030022}{DOI:10.3390/metrology4030022}

\bibitem{Dieks}
D.\ Dieks, \textit{Perspectival quantum realism}, Foundations of Physics \textbf{52}(4), 95 (2022). \href{https://doi.org/10.1007/s10701-022-00611-8}{DOI:10.1007/s10701-022-00611-8 }

\bibitem{Fuchs}
C.\ A.\ Fuchs, \textit{Notwithstanding Bohr, the Reasons for QBism}, Mind Matter, \textbf{15}, 245-–300 (2017). \href{https://doi.org/10.48550/arXiv.1705.03483}{DOI:10.48550/arXiv.1705.03483}

\bibitem{Rovelli}
C.\ Rovelli, \textit{Relational quantum mechanics}, Int. J. Theor. Phys., \textbf{35}, 1637-–1678 (1996). \href{https://doi.org/10.1007/BF02302261}{DOI:10.1007/BF02302261}

\bibitem{DiBiagio}
A.\ Di Biagio, and C.\ Rovelli, \textit{Stable facts, relative facts}, Foundations of Physics \textbf{51}, 1--13 (2021). \href{https://doi.org/10.1007/s10701-021-00429-w}{DOI:10.1007/s10701-021-00429-w }

\bibitem{Healey}
R.\ Healey, \textit{Securing the objectivity of relative facts in the quantum world}, Found.\ Phys \textbf{52}, 88 (2022). \href{https://doi.org/10.1007/s10701-022-00603-8}{DOI:10.1007/s10701-022-00603-8}

\bibitem{Ormrod2}
N.\ Ormrod and J.\ Barrett, \textit{Quantum influences and event relativity}, arXiv:2401.18005 (2024). \href{https://doi.org/10.48550/arXiv.2401.18005}{DOI:10.48550/arXiv.2401.18005}

\bibitem{Mueller}
M.\ P.\ M\"uller, \textit{Law without law: from observer states ot physics via algorithmic information theory}, Quantum \textbf{4}, 301 (2020). \href{https://doi.org/10.22331/q-2020-07-20-301}{DOI:10.22331/q-2020-07-20-301}

\bibitem{Sagona-Stophel}
S.\ Sagona-Stophel, \textit{Falsifiable Tests for Theories that Govern How an Individual's Conscious Experience Traverses Everett's ``Many-Worlds'' Multiverse}, arXiv:2303.08820 (2023). \href{https://doi.org/10.48550/arXiv.2303.08820}{DOI:10.48550/arXiv.2303.08820}

\bibitem{Renner2018}
R.\ Renner, personal communication (2018).

\bibitem{Renes}
J.\ M.\ Renes, \textit{Consistency in the description of quantum measurement: Quantum theory can consistently describe the use of itself}, arXiv:2107.02193 (2021). \href{https://doi.org/10.48550/arXiv.2107.02193}{DOI:10.48550/arXiv.2107.02193}

\bibitem{Fines-theorem}
A.\ Fine,  \textit{Hidden variables, joint probability, and the Bell inequalities}, Phys.\ Rev.\ Lett., \textbf{48}(5), 291 (1982). \href{https://doi.org/10.1103/PhysRevLett.48.291}{DOI:10.1103/PhysRevLett.48.291}

\bibitem{Scarani}
V.\ Scarani, \textit{Bell Nonlocality}, Oxford University Press, Oxford, 2019. \href{https://doi.org/10.1093/oso/9780198788416.001.0001}{DOI:10.1093/oso/9780198788416.001.0001}

\bibitem{Utreras-Alarcon}
A.\ Utreras-Alcarc\'+on, E.\ G.\ Cavalcanti, and H.\ M.\ Wiseman, \textit{Allowing Wigner's friend to sequentially measure incompatible observables}, Proc.\ R.\ Soc.\ A: Math.\ Phys.\ Eng.\ \textbf{480}, 20240040 (2024). \href{https://doi.org/10.1098/rspa.2024.0040}{DOI:10.1098/rspa.2024.0040}

\bibitem{Cavalcanti}
E.\ G.\ Cavalcanti, and H.\ M.\ Wiseman, \textit{Implications of local friendliness violation for quantum causality}, Entropy \textbf{23}(8), 925 (2021). \href{https://doi.org/10.3390/e23080925}{DOI:10.3390/e23080925}

\bibitem{Walleghem}
L.\ Walleghem and R.\ Wagner, \textit{Extended Wigner's friend paradoxes do not require nonlocal correlations}, Phys.\ Rev.\ A \textbf{112}, 022212 (2025). \href{https://doi.org/https://doi.org/10.1103/n4hv-rlgj}{DOI:10.1103/n4hv-rlgj}

\bibitem{Szangolies}
J.\ Szangolies, \textit{The Quantum Rashomon Effect: A Strengthened Frauchiger-Renner Argument}, arXiv:2011.12716 (2023). \href{https://doi.org/10.48550/arXiv.2011.12716}{DOI:10.48550/arXiv.2011.12716}

\bibitem{Carroll}
S.\ M.\ Carroll, \textit{Why Boltzmann brains are bad}, Current controversies in philosophy of science. Routledge, 7--20 (2020). \href{https://doi.org/10.4324/9781315713151-3}{DOI:10.4324/9781315713151-3}

\bibitem{McQueenVaidman}
K.\ J.\ McQueen and L.\ Vaidman, \textit{In defence of the self-location uncertainty account of probability in the many-worlds interpretation}, Stud.\ Hist.\ Philos.\ Mod.\ Phys.\ \textbf{66}, 14--23 (2019). \href{https://doi.org/10.1016/j.shpsb.2018.10.003}{DOI:10.1016/j.shpsb.2018.10.003}

\bibitem{LiVitanyi}
M.\ Li and P.\ Vit\'anyi, \textit{An Introduction to Kolmogorov Complexity and Its Applications}, 3rd edition, Springer (2008). \href{https://doi.org/10.1007/978-0-387-49820-1}{DOI:10.1007/978-0-387-49820-1}

\bibitem{Nagel}
T.\ Nagel, \textit{What is it like to be a bat?}, The Philosophical Review \textbf{83}(4), 435--450 (1974). \href{https://doi.org/10.2307/2183914}{DOI:10.2307/2183914}

\bibitem{Mueller2024}
M.\ P.\ M\"uller, \textit{Algorithmic idealism: what should you believe to experience next?}, Found.\ Phys.\ \textbf{56}, 11 (2026). \href{https://doi.org/10.1007/s10701-026-00913-1}{DOI:10.1007/s10701-026-00913-1}

\bibitem{MyrvoldStatMech}
W.\ C.\ Myrvold, \textit{Probabilities in Statistical Mechanics}, in C.\ Hitchcock and A.\ H\'ajek (eds.), \textit{The Oxford Handbook of Probability and Philosophy}, Oxford University Press (2016). \href{https://doi.org/10.1093/oxfordhb/9780199607617.013.26}{DOI:10.1093/oxfordhb/9780199607617.013.26}

\bibitem{Adlam}
E.\ Adlam, \textit{Against Self-Location}, Br.\ J.\ Philos.\ Sci.\ (2024). \href{https://doi.org/10.1086/732908}{DOI:10.1086/732908}

\bibitem{Kent2}
A.\ Kent, \textit{Quantum reality via late-time photodetection}, Phys.\ Rev.\ A \textbf{96}, 062121 (2017). \href{https://doi.org/10.1103/PhysRevA.96.062121}{DOI:10.1103/PhysRevA.96.062121}

\bibitem{RaymondRobichaud}
G.\ Brassard and P.\ Raymond-Robichaud, \textit{Parallel Lives: A Local-Realistic Interpretation of ``Nonlocal'' Boxes}, Entropy \textbf{21}(1), 87 (2019). \href{https://doi.org/10.3390/e21010087}{DOI:10.3390/e21010087}

\bibitem{Harrigan}
N.\ Harrigan and R.\ W.\ Spekkens, \textit{Einstein, Incompleteness, and the Epistemic View of Quantum States}, Found.\ Phys.\ \textbf{40}, 125--157 (2010). \href{https://doi.org/10.1007/s10701-009-9347-0}{DOI:10.1007/s10701-009-9347-0}

\bibitem{Popescu}
S.\ Popescu, and D.\ Rohrlich, \textit{Causality and non-locality as axioms for quantum mechanics}, In: Hunter, G., Jeffers, S., Vigier, JP. (eds) Causality and Locality in Modern Physics. Fundamental Theories of Physics, vol 97. Springer, Dordrecht. \href{https://doi.org/10.1007/978-94-017-0990-3_45}{DOI:10.1007/978-94-017-0990-3\_45}

\bibitem{DutchBook:Stanford}
S.\ Vineberg, \textit{Dutch Book Arguments}, The Stanford Encyclopedia of Philosophy (Fall 2022 Edition), Edward N. Zalta \& Uri Nodelman (eds.). \url{https://plato.stanford.edu/archives/fall2022/entries/dutch-book/}

\bibitem{GreavesMyrvold}
H.\ Greaves and W.\ Myrvold, \textit{Everett and Evidence}, in S.\ Saunders, J.\ Barrett, A.\ Kent, and D.\ Wallace (eds.), \textit{Many Worlds? Everett, Quantum Theory \& Reality}, Oxford University Press (2010). \href{https://doi.org/10.1093/acprof:oso/9780199560561.003.0011}{DOI:10.1093/acprof:oso/9780199560561.003.0011}

\bibitem{Eddington}
A.\ S.\ Eddington, \textit{The End of the World (From the Standpoint of Mathematical Physics.)}, The Mathematical Gazette \textbf{15}(212), 316--324 (1931). \href{https://doi.org/10.2307/3606671}{DOI:10.2307/3606671 }

\bibitem{Albrecht}
A.\ Albrecht, \textit{Cosmic inflation and the arrow of time}, Science and ultimate reality: Quantum theory, cosmology and complexity, 363--401 (2004). \href{https://doi.org/10.1017/cbo9780511814990.021}{DOI:10.1017/cbo9780511814990.021 }

\bibitem{AlbrechtSorbo}
A.\ Albrecht, and L.\ Sorbo, \textit{Can the universe afford inflation?}, Physical Review D \textbf{70}(6), 063528 (2004). \href{https://doi.org/10.1103/PhysRevD.70.063528}{DOI:10.1103/PhysRevD.70.063528}

\bibitem{Norton}
J.\ D.\ Norton, \textit{You are not a Boltzmann brain}, PhilSci-Archive preprint, item ID 17689 (2015). \url{http://philsci-archive.pitt.edu/id/eprint/17689}


\end{thebibliography}
\end{document}